\documentclass[12pt]{article}
\usepackage{epsf,epsfig,amssymb,amsmath,graphicx,float,latexsym} 

\newcommand{\lsim}{\raisebox{-0.13cm}{~\shortstack{$<$ \\[-0.07cm] $\sim$}}~}
\newcommand{\gsim}{\raisebox{-0.13cm}{~\shortstack{$>$ \\[-0.07cm] $\sim$}}~}

\begin{document}
\renewcommand{\thefootnote}{\fnsymbol{footnote}}

\begin{titlepage}
  
\begin{flushright}
KIAS--P06010 \\
ICRR-Report--524--2005--7 \\
March 2006
\end{flushright}

\begin{center}

\vspace{1cm}

{\Large {\bf Abundance of Cosmological Relics in Low--Temperature
    Scenarios }}

\vspace{1cm}

{\bf Manuel Drees}$^{a,b,\,}$\footnote{drees@th.physik.uni-bonn.de},
{\bf Hoernisa Iminniyaz}$^{a,c,\,}$\footnote{hoernisa@th.physik.uni-bonn.de},
{\bf Mitsuru Kakizaki}$^{a,d,\,}$\footnote{kakizaki@th.physik.uni-bonn.de} \\

\vskip 0.15in
{\it
$^a${Physikalisches Institut der Universit\"at Bonn,
Nussallee 12, 53115 Bonn, Germany}\\
$^b$Korea Institute of Advanced Studies, School of Physics, Seoul 130--012,
South Korea 
\\
$^c${Physics Dept., Univ. of Xinjiang, 830046 Urumqi, P.R. China}\\
$^d${ICRR, Univ. of Tokyo, Kashiwa 277-8582, Japan }\\
}
\vskip 0.5in

\abstract{We investigate the relic density $n_\chi$ of non--relativistic
  long--lived or stable particles $\chi$ in cosmological scenarios in which
  the temperature $T$ is too low for $\chi$ to achieve full chemical
  equilibrium. The case with a heavier particle decaying into $\chi$ is also
  investigated. We derive approximate solutions for $n_\chi(T)$ which
  accurately reproduce numerical results when full thermal equilibrium is not
  achieved. If full equilibrium is reached, our ansatz no longer reproduces
  the correct temperature dependence of the $\chi$ number density. However, it
  does give the correct final relic density, to an accuracy of about 3\% or
  better, for {\em all} cross sections and initial temperatures.}

\end{center}
\end{titlepage}
\setcounter{footnote}{0}

\section{Introduction}

The production of massive, long--lived or stable relic particles $\chi$ plays
a crucial role in particle cosmology \cite{kotu}. The perhaps most important
example is the production of Massive Weakly--Interacting Particles (WIMPs),
which may constitute most of the Dark Matter in the universe \cite{silkrev}.
Alternatively, WIMPs may only be meta--stable, and decay into even more weakly
interacting particles (e.g. gravitinos or axinos) that form the Dark Matter
\cite{superwimp}. Even if WIMP decays do not produce Dark Matter particles,
the WIMP density is tightly constrained by analyses of Big Bang
Nucleosynthesis (BBN)\cite{bbnrev}.

It is usually assumed that the WIMPs were in full thermal and chemical
equilibrium in the radiation--dominated epoch after the period of last entropy
production, which in standard cosmology means after the end of inflation. In
this ``standard'' scenario the $\chi$ number density $n_\chi(T)$ drops
exponentially once the temperature $T$ falls below the mass $m_\chi$ of the
relic particles, until the freeze--out temperature $T_F$ is reached, where the
production of $\chi$ particles from the thermal bath becomes negligible. In
this case accurate semi--analytical expressions for $n_\chi(T \ll T_F)$ have
been derived \cite{standard,except}; one finds that the $\chi$ relic density
is essentially inversely proportional to the thermal average of the effective
$\chi$ annihilation cross section into lighter particles. The case of
additional late entropy production, at $T \ll T_F$, can also be treated
analytically, by multiplying the standard result with a ``dilution factor''
due to the late--produced entropy \cite{late}.

For typical WIMP scenarios, $T_F \simeq m_\chi/20$. The standard treatment can
work only if the maximal temperature after inflation, usually called the
reheat temperature $T_R$, is (much) larger than $T_F$. The assumption $T_R \gg
T_F$ is not implausible, since the scale of inflation has to be quite high,
typically $\sim 10^{13}$ GeV in simple models, in order to achieve the right
order of magnitude of density perturbations \cite{inflatrev}. On the other
hand, we have direct observational evidence (from BBN) only for temperatures
$T \lsim $ (few) MeV \cite{trmin1,trmin2}, which is well below $T_F$ for most
current WIMP candidates \cite{silkrev}. It is therefore legitimate to
investigate scenarios with $T_R \lsim T_F$ \cite{low,low_others,gg}.

We should emphasize at this point that $T_R$ may not have been the highest
temperature of the thermal plasma after inflation: given sufficiently fast
thermalization, the inflaton decay products can attain a temperature $T_{\rm
  max} \gg T_R$ while the total energy density of the universe is still
dominated by inflatons \cite{kotu}. $\chi$ particles may therefore have been
in thermal equilibrium for some range of temperatures $T > T_R$
\cite{ckr,trmin1,ad,low,decay}, even if they never were in equilibrium in the
radiation--dominated epoch. However, an analytical treatment of the
re--heating epoch where $T > T_R$ was possible faces several complications not
present in the radiation--dominated epoch: the entropy density was not
constant, non--perturbative (and non--exponential) inflaton decays might have
been important \cite{preheat}, and there might have been significant
non--thermal sources of $\chi$ particles \cite{ad2,ad,decay}. On the other
hand, in supersymmetric scenarios thermalization of the inflaton decay
products might be delayed by large vacuum expectation values of scalar fields
along flat directions of the potential \cite{delay}. In this paper we evade
these complications by treating the $\chi$ number density at some initial
temperature $T_0$ as a free parameter; in the absence of late entropy
production, $T_0$ should be close to the reheat temperature $T_R$ (depending
on the exact definition of $T_R$).

Existing treatments of thermal WIMP production
\cite{standard,except,ckr,trmin1,ad,low,decay} assume that $n_\chi$ had either
achieved full equilibrium, or was completely out of equilibrium (i.e.,
annihilation of $\chi$ particles was always negligible).  As already noted, in
the former case one finds that the relic density is inversely proportional to
the thermal average of the $\chi$ annihilation cross section. Not
surprisingly, if $\chi$ annihilation can be neglected, one finds that the
contribution to the $\chi$ relic density from thermal production is directly
proportional to this cross section. Here we provide an approximate analytic
treatment that also works in the intermediate region, where (for some range of
temperatures) both thermal production and annihilation of $\chi$ particles
were important. It is based on an expansion in the effective annihilation
cross section. To leading order, only the production term is kept in the
Boltzmann equation describing the evolution of $n_\chi(T)$; this corresponds
to the ``completely out of equilibrium'' scenario analyzed previously. The
first correction includes $\chi$ annihilation, treating it as a small
perturbation. This still allows an analytic solution, in terms of the
exponential integral of first order ${\rm E}_1$, which we only need for large
values of its argument. If $n_\chi(T_0) =0$, the first--order result is linear
in the annihilation cross section $\sigma$, while the correction is ${\cal
  O}(\sigma^3)$. Our most important, and (to us) rather surprising, result is
that terms of higher order in $\sigma$ can be ``re--summed'' using a simple
trick. This can be shown to be exact in the simple case where $n_\chi(T_0) >
0$ and thermal production of $\chi$ particles is negligible\footnote{In this
  case the leading order result is trivial, i.e.  ${\cal O}(\sigma^0)$, while
  the first correction is ${\cal O}(\sigma)$.}, and works numerically also for
non--negligible thermal production. In fact, for $T \ll T_0$ our formulae
reproduce the exact numerical results to 3\% or better even for combinations
of parameters where $n_\chi$ achieved complete equilibrium, i.e. our new
formulae are also accurate in scenarios where the ``standard'' result
\cite{standard} is applicable.

The outline of our paper is as follows. In Sec.~2 we briefly review the
calculation of the relic abundance in the ``standard'' scenario, where it is
assumed that the relic particles attained full thermal equilibrium. In
Sec.~3 we will discuss our analytic calculation of the $\chi$ relic
abundance in scenarios where the temperature was too low for $\chi$ particles
to have been in full equilibrium.  In Sec.~4 we apply this method to
more complicated scenarios, which include non--thermal $\chi$ production from
the decay of a heavier particle, still assuming the universe to be radiation
dominated. Finally, Sec.~5 is devoted to a brief summary and some
conclusions, while some technical details are given in the Appendix.

\section{Relic Abundance in the Standard Cosmological Scenario} 

We briefly review the calculation of the relic density of long--lived or
stable particles $\chi$ in the standard cosmological scenario \cite{standard},
which assumes that the relic particles were in thermal equilibrium in the
early universe and decoupled when they were non--relativistic. The relic
density can be calculated by solving the Boltzmann equation which describes
the time evolution of the number density $n_\chi$ in the expanding universe
\cite{kotu},
\begin{equation}
  \frac{dn_{\chi}}{dt} + 3 H n_{\chi} =  - \langle \sigma v \rangle
  (n^2_{\chi} - n_{\chi,{\rm eq}}^2)~,
  \label{eq:boltzmann_n}
\end{equation}
with $n_{\chi, {\rm eq}}$ being the equilibrium number density of the relic
particles, $H$ the Hubble parameter and $\langle \sigma v \rangle$ the thermal
average of the annihilation cross section $\sigma$ multiplied with the
relative velocity $v$ of the two annihilating $\chi$ particles. The first
(second) term on the right--hand side (rhs) of Eq.(\ref{eq:boltzmann_n})
describes the decrease (increase) of the number density due to annihilation
into (production from) lighter particles. The equilibrium density in the
non-relativistic limit is given by
\begin{equation} \label{eq:eq}
  n_{\chi,{\rm eq}} = g_\chi ~{\left( \frac{m_\chi T}{2 \pi} \right)}^{3/2}
  {\rm e}^{-m_\chi/T}~, 
\end{equation}
where $m_\chi$ and $g_\chi$ are the mass and the number of internal degrees of
freedom of $\chi$, respectively. In the standard cosmological scenario, it is
assumed that $\chi$ was in thermal equilibrium for $T \gsim m_\chi$.  In other
words, $\chi$ rapidly annihilated with its own antiparticle into lighter
states and vice versa. At later times $T \ll m_\chi$, the annihilation rate
$\Gamma_{\chi} = n_{\chi} \langle \sigma v \rangle $ dropped below the
expansion rate $H$. Therefore $\chi$ particles were no longer able to
annihilate efficiently and the number density per co--moving volume became
constant.  The temperature at which the particle decouples from the thermal
bath is called freeze--out temperature $T_F$.

The Boltzmann equation~(\ref{eq:boltzmann_n}) can be rewritten by introducing
the new variables $Y_\chi = n_\chi/s$ and $Y_{\chi,{\rm eq}} = n_{\chi,{\rm
    eq}}/s$, where the entropy density $s = (2 \pi^2/45) g_* T^3$ with $g_*$
being the number of the relativistic degrees of freedom.  Assuming that the
universe expands adiabatically, the entropy per comoving volume is conserved.
Hence we obtain\footnote{Here we assume $\dot{g}_*=0$. This is usually
  justified since, as we will see below, $n_\chi$ has non--trivial time
  dependence only for a rather narrow range of temperatures; moreover, except
  during the QCD phase transition at $T \simeq 200$ MeV, $g_*$ changes slowly,
  i.e. $d g_* / d x \ll g_*$.} $\dot{n}_\chi + 3 H n_\chi = s \dot{Y}_\chi$.
In the radiation dominated era the Hubble parameter is given by
\begin{equation} \label{eq:bx}
  H = \frac{\pi T^2}{M_{\rm Pl}} \sqrt{\frac{g_*}{90}}\, , 
  \quad t = \frac{1}{2 H}\, ,
\end{equation}
where $M_{\rm Pl}$ is the reduced Planck mass, $M_{\rm Pl} = 2.4 \times
10^{18}$~GeV.  By introducing the inverse scaled temperature $x = m/T$, the
Boltzmann equation~(\ref{eq:boltzmann_n}) becomes
\begin{eqnarray} \label{eq:boltzmann}
  \frac{dY_\chi}{dx} = - 1.32~ m_\chi M_{\rm Pl} \sqrt{g_*} \langle \sigma v
  \rangle x^{-2} (Y_\chi^2 - Y_{\chi,{\rm eq}}^2)\, .
\end{eqnarray}
In most (although not all \cite{except}) cases the cross section is well
approximated by a non--relativistic expansion:
\begin{equation} \label{eq:ab}
\langle \sigma v \rangle = a + b \langle v^2 \rangle
  + {\cal O}(\langle v^4 \rangle) =  a + 6 b/x + {\cal O}(1/x^2)\, .
\end{equation}
Here $a$ is the $v \rightarrow 0$ limit of the contribution to $\sigma v$
where the two annihilating $\chi$ particles are in an $S$ wave. If $S$ wave
annihilation is suppressed, $b$ describes the $P$ wave contribution to $\sigma
v$. In the following we treat $a$ and $b$ as free parameters. In terms of the
variable $\Delta = Y_\chi - Y_{\chi,{\rm eq}}$, the Boltzmann
equation~(\ref{eq:boltzmann}) can be rewritten as
\begin{equation}
  \frac{d \Delta}{dx} = - \frac{d Y_{\chi, {\rm eq}}}{dx} - \lambda \Delta
  (2 Y_{\chi,{\rm eq}} + \Delta)\, , 
  \label{eq:delta}
\end{equation}
where
\begin{equation}
  \lambda = 1.32~m_\chi M_{\rm Pl} \sqrt{g_*} (a + 6 b/x)~x^{-2}\, .
\end{equation}
An analytic solution can be obtained by considering the equation in two
extreme regimes. At early times ($x \ll x_F$), $Y$ tracks its equilibrium
value $Y_{\rm eq}$ very closely. Therefore $\Delta$ and $d\Delta/dx$ are
small. Ignoring $\Delta^2$ and $d\Delta/dx$, we obtain
\begin{eqnarray} \label{eq:early}
  \Delta \simeq \frac{1}{2 \lambda }\, ,
\end{eqnarray}
where we used $d Y_{\chi,{\rm eq}}/dx \simeq - Y_{\chi,{\rm eq}}$ for $x \gg
1$. At late times ($x \gg x_F$), one can ignore the production term in the
Boltzmann equation:
\begin{equation} 
  \frac{d \Delta}{dx} \simeq - \lambda \Delta^2\, .
\end{equation}
Integrating this equation from $x_F$ to infinity and using the fact that
$\Delta(x_F) \gg \Delta(\infty)$, we have
\begin{equation}
  Y_{\chi, \infty} \equiv Y_\chi(x \gg x_F) = \frac{x_F}{1.32~m_\chi M_{\rm Pl}
  \sqrt{g_*(x_F)} (a + 3 b/x_F)}\, .  
\end{equation}
It is useful to express the energy density as $\Omega_{\chi} = \rho_{\chi} /
\rho_c$, where $\rho_c = 3 H^2_0 M^2_{\rm Pl} $ is the critical density of the
universe. The present energy density of the relic particle is given by
$\rho_{\chi} = m_\chi n_{\chi, \infty} = m_\chi s_0 Y_{\chi,\infty}$, with
$s_0 \simeq 2900~ {\rm cm}^{-3}$ being the present entropy density. Finally,
we obtain the standard approximate formula for the relic density:
\begin{equation} \label{eq:omold}
  \Omega_{\chi} h^2 \simeq \frac{8.7 \times 10^{-11}~x_F~{\rm
  GeV}^{-2}}{\sqrt{g_* (x_F)} (a + 3 b/x_F)}\, ,
\end{equation}
where $h$ is the scaled Hubble constant, $h \simeq 0.7$. Notice that the relic
density of the particle is inversely proportional to the annihilation cross
section and that there is no explicit dependence on the mass of the particle.
Calculating the cross section and the freeze-out temperature is sufficient for
predicting the relic density.  Freeze-out occurs when the deviation $\Delta$
is of the same order as the equilibrium value:
\begin{equation}
  \Delta (x_F) = \xi Y_{\chi,{\rm eq}}(x_F)\, ,
\end{equation}
where $\xi$ is a numerical constant of order unity.  Substituting the early
time solution of Eq.(\ref{eq:early}) into this equation, $x_F$ is obtained by
iteratively solving
\begin{equation} \label{eq:x_F}
  x_F = \ln \frac{0.382\, \xi m_\chi M_{\rm Pl} g_\chi (a + 6 b/x_F)}{\sqrt{x_F
      g_*(x_F)} }\, .
\end{equation}
It is known that the choice $\xi = \sqrt{2} - 1$ gives a good approximation of
exact numerical results for the relic density (\ref{eq:omold}). The decoupling
temperature depends only logarithmically on the cross section. For WIMPs, we
typically obtain $x_F \simeq 22$.

\section{Relic Abundance in a Low--Temperature Scenario}
\setcounter{footnote}{0}

Eq.(\ref{eq:omold}) implies that the relic density predicted in the standard
cosmological scenario, in which $\chi$ particles are assumed to have been in
full equilibrium, would be quite high unless the cross section is as large
as\footnote{We use natural units, where $\hbar = c = k_B = 1$, so that both
  $\sigma$ and $\sigma v$ have dimensions GeV$^{-2}$. Numerically, $10^{-9}$
  GeV$^{-2} = 0.388$ pb $= 1.16 \cdot 10^{-26}$ cm$^3$/s.} $\sim 10^{-9}$
GeV$^{-2}$.  Bearing this situation in mind, it is important to explore
scenarios where the relic density comes out smaller than the standard
calculation and find a useful formula which properly describes the behavior of
the relic abundance.

For later convenience we first rewrite the Boltzmann equation
(\ref{eq:boltzmann}), using Eq.(\ref{eq:eq}): 
\begin{eqnarray} \label{eq:bx1}
  \frac{dY_\chi}{dx} = - f \left( a + \frac{6b}{x} \right) \frac{1}{x^2}
  \left( Y_\chi^2 - c x^3 {\rm e}^{-2x} \right)\, ,
\end{eqnarray}
where
\begin{eqnarray}
  f = 1.32\, \sqrt{g_*} m_\chi M_{\rm Pl}~, \quad c = 0.0210\, g_\chi^2/g_*^2
\end{eqnarray}
are constants.  Eqs.(\ref{eq:boltzmann}) and (\ref{eq:bx1}) assume that $\chi$
remains in kinetic equilibrium through the entire period with non--negligible
time dependence of $Y_\chi$. This is reasonable, since kinetic equilibrium can
be maintained through elastic scattering of $\chi$ particles on particles in
the thermal plasma. The rate for such reactions exceeds the $\chi$
annihilation rate by a factor $\propto Y_\chi^{-1} \gsim 10^7$ for
temperatures of interest.  For our numerical examples, we consider a Majorana
fermion with $m_\chi= 100$ GeV and $g_\chi = 2$ as the relic particle. We
choose the relativistic degrees of freedom to be $g_* = 90$; this approximates
the prediction of the Standard Model of particle physics for temperatures
around $10$ GeV.

\begin{figure}[t]
  \begin{center}
    \hspace*{-0.5cm}
    \scalebox{0.63}{\includegraphics*{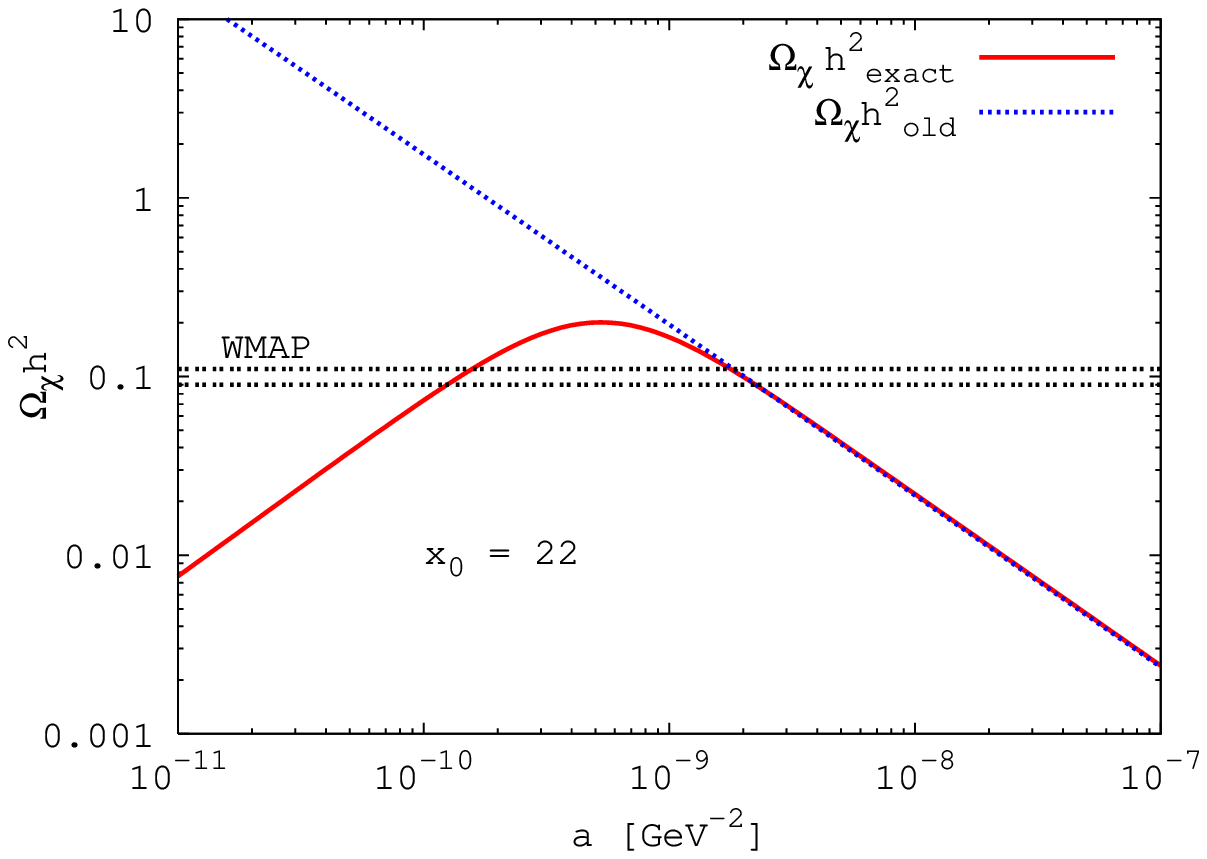}}
    \put(-115,-12){(a)}
    \hspace*{-0.5cm}
    \scalebox{0.63}{\includegraphics*{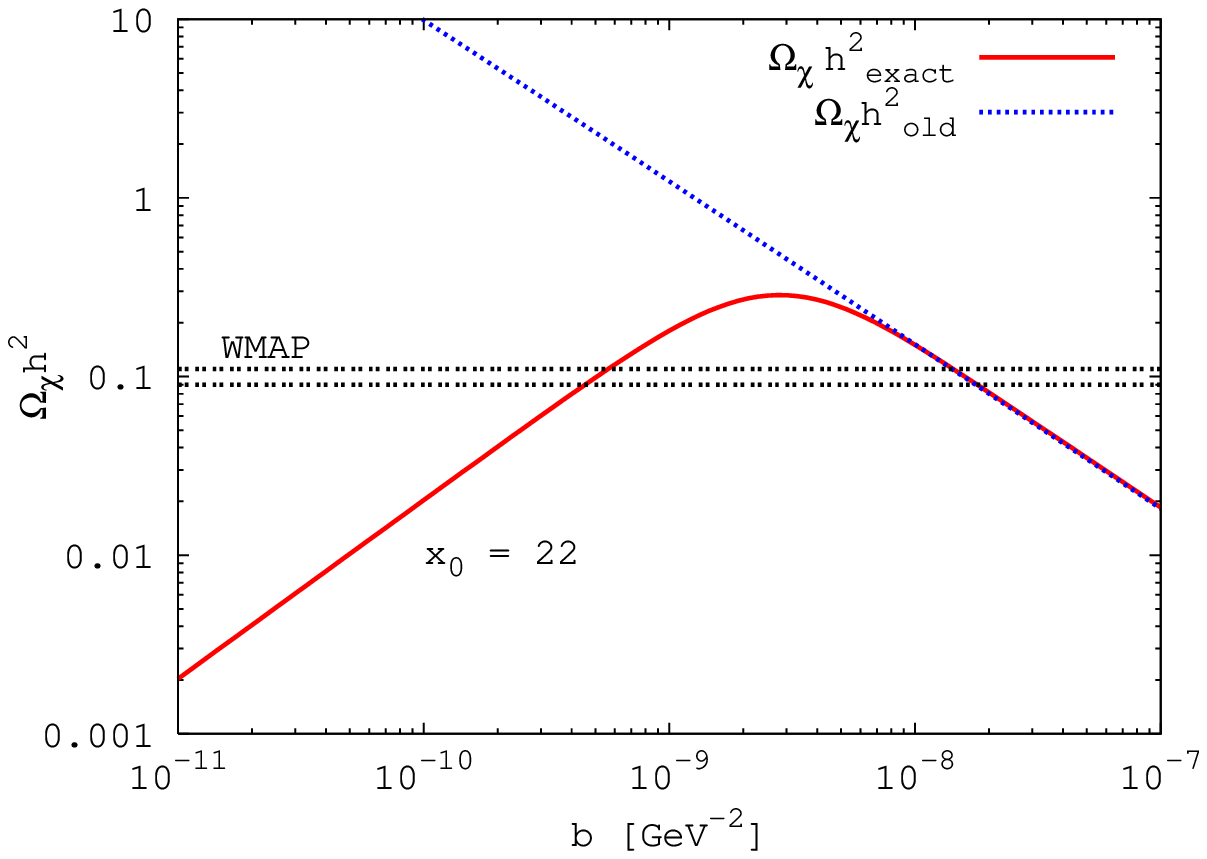}}
    \put(-115,-12){(b)}
\caption{\footnotesize Predicted present relic density $\Omega_\chi h^2$ as
  function of the $a$ and $b$ contributions to the total cross section, see
  Eq.(\ref{eq:ab}); in frame (a), $b=0$ whereas in (b), $a=0$.  We consider
  two extreme cases: $\chi$ particles were in full thermal equilibrium (dotted
  blue line) or the number density of $\chi$ vanished (solid red line) at $x_0
  = 22$.  The two horizontal double--dotted black lines correspond to the
  1$\sigma$ upper and lower bounds of the dark matter abundance \cite{wmap}.}
\label{fig:omegah2}
\end{center}
\end{figure}

Figure~\ref{fig:omegah2} shows that the relic density can be reduced if the
particles never reach thermal equilibrium because of the low reheat
temperature after inflation.  The solid red curves depict the predicted
present relic density $\Omega_\chi h^2$ as function of $a$ (a) and $b$ (b)
defined in Eq.(\ref{eq:ab}). Here we assume that the relic abundance vanished
at the initial temperature of $x_0 = 22$, which is around the typical WIMP
decoupling temperature. Here, as well as in the subsequent figures, the
``exact'' numerical solution of the Boltzmann equation (\ref{eq:bx1}) has been
obtained using the Runge--Kutta algorithm, with a step size that increases
quickly with increasing $x-x_0$. For large cross section we observe
$\Omega_\chi h^2 \propto 1/\langle \sigma v\rangle$, in accord with the
``standard'' prediction (\ref{eq:omold}). However, when the cross section is
reduced, the relic density reaches a maximum, and then decreases $\propto
\langle \sigma v\rangle$. For the given choice of initial conditions, there
are therefore two distinct ranges in $\langle \sigma v\rangle$ where the relic
density comes out in the desired range \cite{wmap}.

In the following we attempt to find a convenient analytic formula applicable
even to low temperature scenarios.  As zeroth order solution of
Eq.(\ref{eq:bx1}) we consider the case where $\chi$ annihilation is completely
negligible,
\begin{eqnarray} \label{eq:e0}
  \frac{d Y_0}{dx} = f c (a x + 6b)\, {\rm e}^{-2x}\, .
\end{eqnarray}
This equation can easily be integrated, giving
\begin{eqnarray} \label{eq:Y_0}
  Y_0(x) = f c \left[ \frac{a}{2} (x_0 {\rm e}^{-2 x_0} - x {\rm e}^{-2 x} )
    + \left( \frac{a}{4} + 3 b \right) ({\rm e}^{-2 x_0} - {\rm e}^{-2 x})
    \right] + Y_\chi(x_0)\, .
\end{eqnarray}
For $ x \gg x_0 $, the relic abundance of the particles becomes constant,
\begin{eqnarray} \label{eq:y0as}
  Y_{0,\infty} \equiv Y_0(x \gg x_0) = f c \left[ \frac{a}{2} x_0 {\rm e}^{-2
      x_0} + \left( \frac{a}{4} + 3 b \right) {\rm e}^{-2 x_0} \right] +
      Y_\chi(x_0)\, . 
\end{eqnarray}
The corresponding prediction for the present relic density is given by
\begin{equation}
 \Omega_{\chi} h^2 = 2.8 \times 10^8~ m_\chi Y_{0,\infty}~ \mbox{GeV}^{-1}\, .
\end{equation}
Notice that the relic density is proportional to the cross section, although
the coefficient of proportionality depends on whether $a$ or $b$ is dominant.

So far no analytic solution has been known for the in--between case where both
annihilation and production play a crucial role in determining the relic
abundance while thermal equilibrium is not fully achieved.  We now attempt to
connect the standard scenario ($T_R>T_F$) and the low reheat temperature
scenario ($T_R<T_F$) using some analytic method.

Since we already have the solution only including the production term, the
most natural extension is to add a correction term which describes the effect
of annihilation on the solution for the pure production case:
\begin{eqnarray} \label{eq:first_order}
  Y_1 = Y_0 + \delta\, .
\end{eqnarray}
By definition $\delta$ vanishes at the initial temperature. Since it describes
the effect of $\chi$ annihilation, it is negative for $x>x_0$. As long as
$|\delta|$ is small compared to $Y_0$, the evolution equation for $\delta$ is
given by
\begin{eqnarray} \label{eq:eq_delta}
  \frac{d \delta}{dx} = - f \left( a + \frac{6b}{x} \right)
  \frac{Y_0(x)^2}{x^2}\, .
\end{eqnarray}
Using Eq.(\ref{eq:Y_0}) for $Y_0(x)$, this can again be integrated:
\begin{eqnarray}
  \delta(x) & = & - f^3 c^2 \left[ \frac{1}{4} a^3 F^4_0(x,x_0)
    + \frac{1}{4} a^2(a+18b) F^4_1(x,x_0) \right.
  \nonumber \\
  && \hspace*{12mm} \left. 
+ \frac{1}{16} a (a + 12 b)(a + 36 b) F^4_2(x,x_0)
    + \frac{3}{8} b (a + 12 b)^2 F^4_3(x,x_0) \right] \nonumber \\
  &+& Y_{0,\infty} f^2 c \left[ a^2 F^2_1(x,x_0)
    + \frac{1}{2} a (a + 24 b) F^2_2(x,x_0)
+ 3 b (a + 12 b) F^2_3(x,x_0) \right] \nonumber \\
  &-& Y_{0,\infty}^2 f \left[ a F^0_2(x,x_0)
      + 6b F^0_3(x,x_0) \right]\, ,
    \label{eq:delta1}
\end{eqnarray}
where
\begin{eqnarray}
  F^m_n(x,x_0) = \int_{x_0}^x dt \ \frac{{\rm e}^{-mt}}{t^n}\, , \quad m=0,2,4\, ,
  \quad n=1,2,3\, .  
\end{eqnarray}
The functions $F^m_n(x,x_0)$ can be expressed analytically in terms of the
exponential integral of first order E$_1(x)$; a complete list of the relevant
$F_n^m$ is given in the Appendix, Eqs.(\ref{eq:app_f}). At late times, $x
\rightarrow \infty$, this simplifies to
\begin{eqnarray} \label{eq:deltaas}
\delta(x \to \infty) & = & - f^3 c^2 {\rm e}^{-4x_0} \left[
  \frac{a^3}{4}x_0 + \frac{a^2 ( a + 60 b )}{16}
  - \frac{9 a b (a - 16b)}{8 x_0} \right. 
\nonumber \\
&& \hspace*{20mm} \left. + \frac{9b (5a^2 - 56ab + 96b^2)}{32x_0^2} \right]~
\nonumber \\
& - & f^2 c {\rm e}^{-2 x_0} Y_\chi(x_0) \left[ a^2 + \frac{9 ab}{x_0} 
  - \frac{9b(a - 4b)}{2 x_0^2} \right]
\nonumber \\
& - & f (Y_\chi(x_0))^2 \left( \frac{a}{x_0} + \frac{3b}{x_0^2} \right) \, ,
\end{eqnarray}
where we omit higher order terms than ${\cal O}(1/x_0^2)$. Notice that we
discard 
${\cal O} (1/x^2)$ and ${\cal O} (1/x^3)$ terms in $\langle \sigma v 
\rangle$, which also contribute to higher order terms in
Eq.(\ref{eq:deltaas}). If $a \neq 0$ we therefore expect additional terms
${\cal O}(1/x_0)$ from terms not included in Eq.(\ref{eq:ab}); if $a=0$,
higher order terms in the expansion of the cross section only contribute at
${\cal O}(1/x_0^3)$ in Eq.(\ref{eq:deltaas}).

Since, for vanishing initial abundance, $Y_0$ is proportional to the cross
section $\sigma$, $\delta$ is proportional to $\sigma^3$. On the other hand,
for sufficiently large cross section we want to recover the standard
expression, where $Y_\chi(x \rightarrow \infty) \propto 1/\langle \sigma v
\rangle$. This suggests to rewrite our ansatz (\ref{eq:first_order}) as
\begin{equation} \label{eq:resum}
Y_1 = Y_0 + \delta = Y_0 \left( 1 + \frac {\delta} {Y_0} \right) \simeq
\frac {Y_0} {1 - \delta/Y_0} \equiv Y_{1,r}\, .
\end{equation}
Although the final approximate equality in Eq.(\ref{eq:resum}) only holds for
$|\delta| \ll Y_0$, we note that the resulting expression has the right
behavior, $Y_{1,r} \propto 1/\sigma$, for large cross section. In the
following we will show that this ``resummation'' of the correction $\delta$
is indeed able to describe the relic density for a wide range of cross
sections and temperatures, including scenarios where the standard treatment is
applicable. 

In fact, this ansatz solves the Boltzmann equation (\ref{eq:bx1}) {\em
  exactly} in the simple case where thermal $\chi$ production can be ignored,
but $Y_\chi(x_0)$ is sizable, leading to significant $\chi$ annihilation. In
this case Eq.(\ref{eq:bx1}) reduces to
\begin{eqnarray} \label{eq:b1}
  \frac{dY_\chi}{dx} = - f \left( a + \frac{6b}{x} \right)
  \frac{Y_\chi^2} {x^2}\, .
\end{eqnarray}
This equation can easily be solved analytically. The solution decreases
monotonically from its initial value $Y_\chi(x_0)$:
\begin{equation}
  Y_\chi = \frac{Y_\chi(x_0)}{1 + f Y_\chi(x_0) \left[ a(1/x_0 - 1/x) +
      3b(1/x_0^2 - 1/x^2) \right] }\, . 
  \label{eq:Y_annihilation}
\end{equation}
In order to treat this case using the formalism of
Eqs.(\ref{eq:e0})--(\ref{eq:resum}), we simply drop all terms which depend
exponentially on $x$ or $x_0$; these terms come from thermal $\chi$
production, and are obviously very small for sufficiently small initial
temperature. The zeroth order solution (\ref{eq:Y_0}) then obviously reduces
to the constant $Y_\chi(x_0)$, and the correction $\delta$ of
Eq.(\ref{eq:delta1}) simplifies to
\begin{eqnarray} \label{eq:check}
\delta(x) &\rightarrow& - f \left( Y_\chi(x_0) \right)^2 \left[ a F_2^0(x,x_0)
  +  6 b F_3^0(x,x_0) \right] \nonumber \\
&=& - f \left( Y_\chi(x_0) \right)^2 \left[ a \left( \frac {1}{x_0} - \frac {1}
  {x} \right) + 3 b \left( \frac {1} {x_0^2} - \frac {1} {x^2} \right) \right]
\, ;
\end{eqnarray}
in the last step we have used the last two Eqs.(\ref{eq:app_f}). Inserting
this in the last expression in Eq.(\ref{eq:resum}), we indeed recover the
exact solution (\ref{eq:Y_annihilation}), as advertised.

In principle, we can add further correction terms to the first order 
approximation of Eq.(\ref{eq:first_order}),
\begin{eqnarray}
  Y_\chi = Y_0 + \delta + \delta_2 + \delta_3 + \cdots\, .
\end{eqnarray}
The above discussion shows that this corresponds to an expansion in powers of
$\langle \sigma v \rangle$. Since $Y_0 > 0$ and $\delta < 0$ by definition, the
systematic expansion will lead to an alternating series which possesses good
convergence properties. However, this type of expansion is quite cumbersome
because $|\delta|$ often dominates over $Y_0$ for not very small cross
sections, as we will explicitly see later. Therefore the re--summed ansatz
$Y_{1,r}$ of Eq.(\ref{eq:resum}) is much more convenient. We will see that it
often provides a good approximation to the exact solution even if thermal
$\chi$ production is not negligible.

\begin{figure}[t!]
  \begin{center}
    \hspace*{-0.5cm} \scalebox{0.63}{\includegraphics*{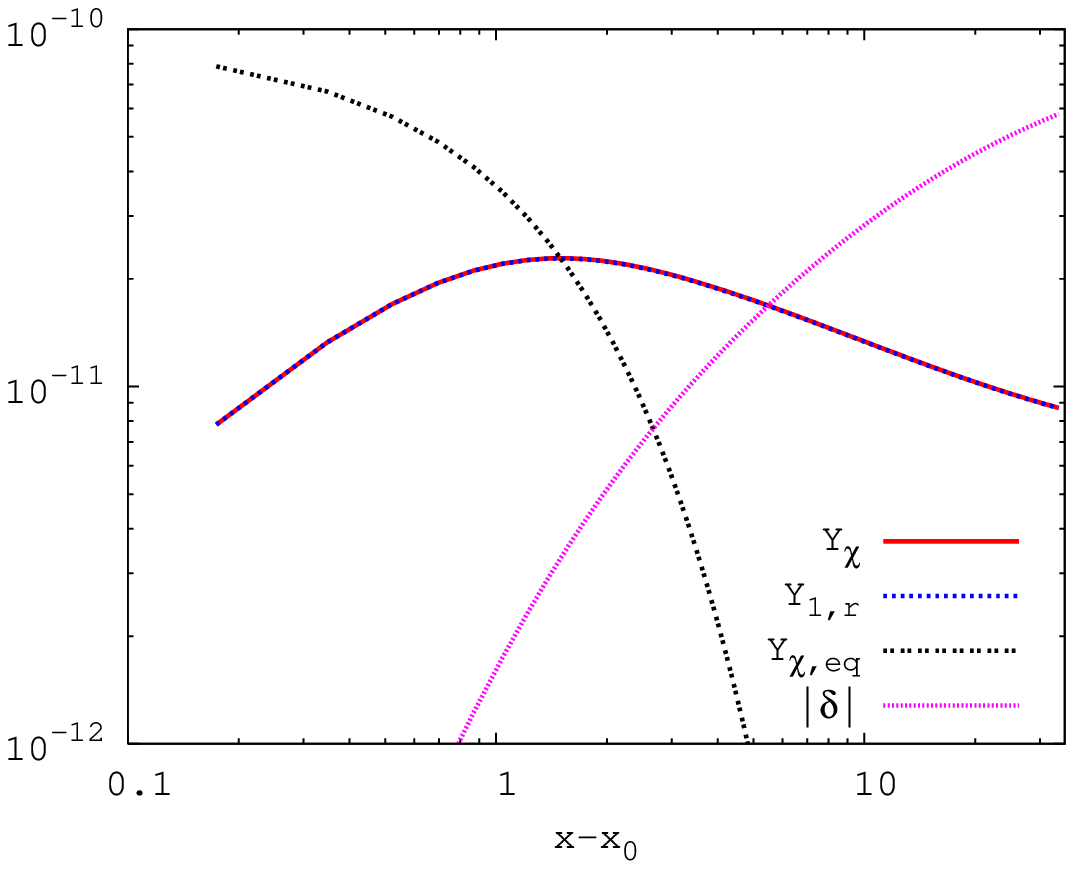}}
    \put(-115,-12){(a)} 
    \hspace*{-0.5cm} \scalebox{0.63}{\includegraphics*{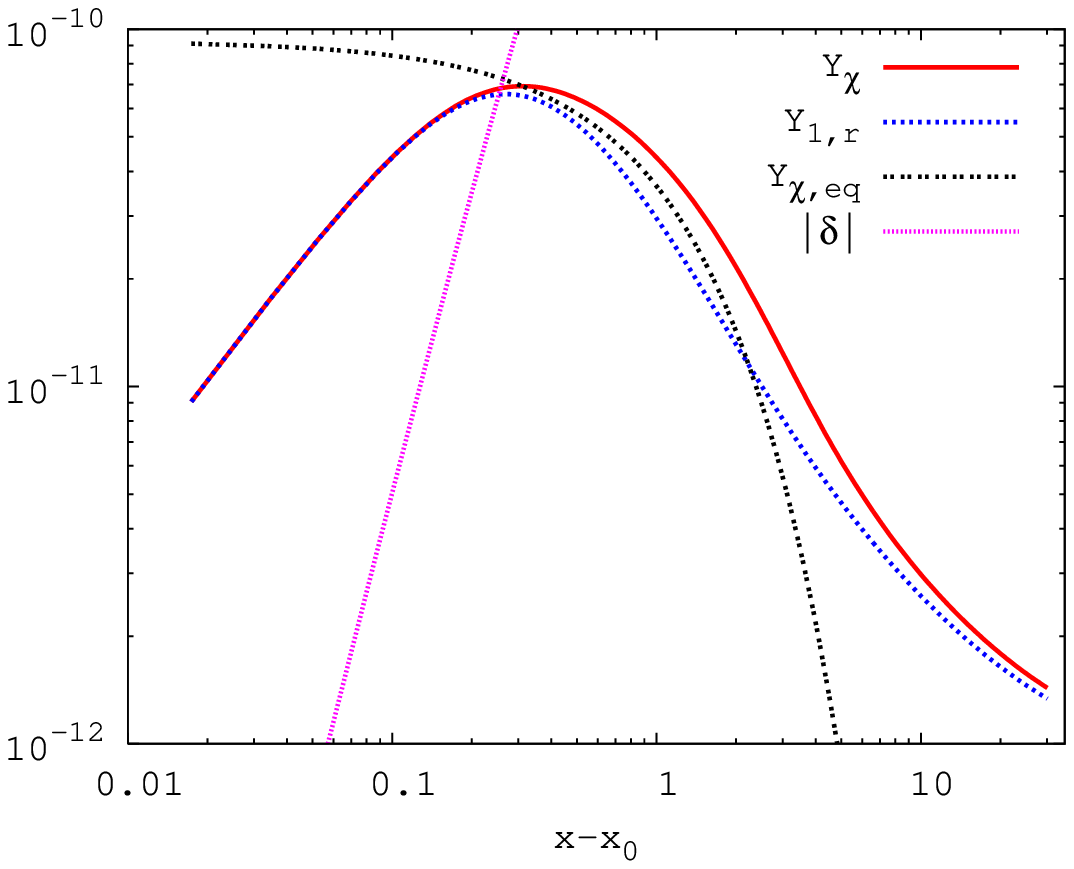}} 
    \put(-115,-12){(b)}
    \vspace{0.5cm} 
    \hspace*{-0.5cm} \scalebox{0.63}{\includegraphics*{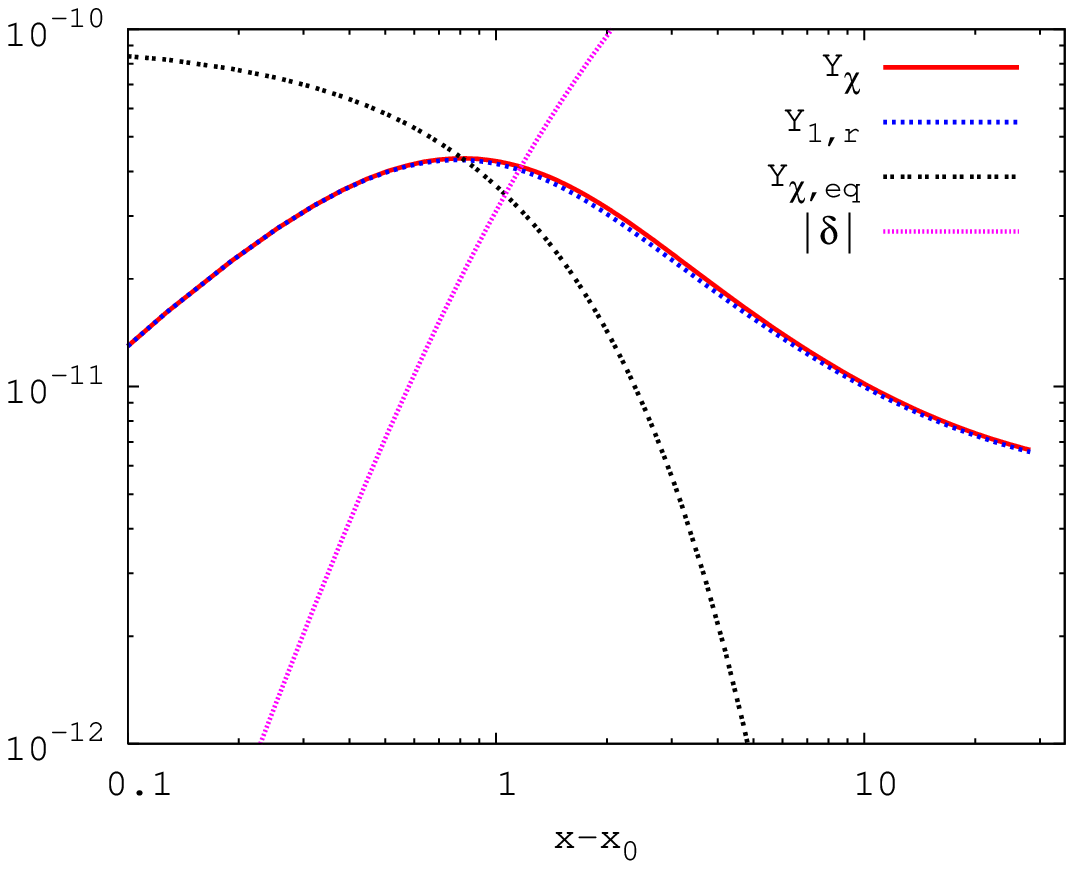}} 
    \put(-115,-12){(c)}
    \hspace*{-0.5cm} \scalebox{0.63}{\includegraphics*{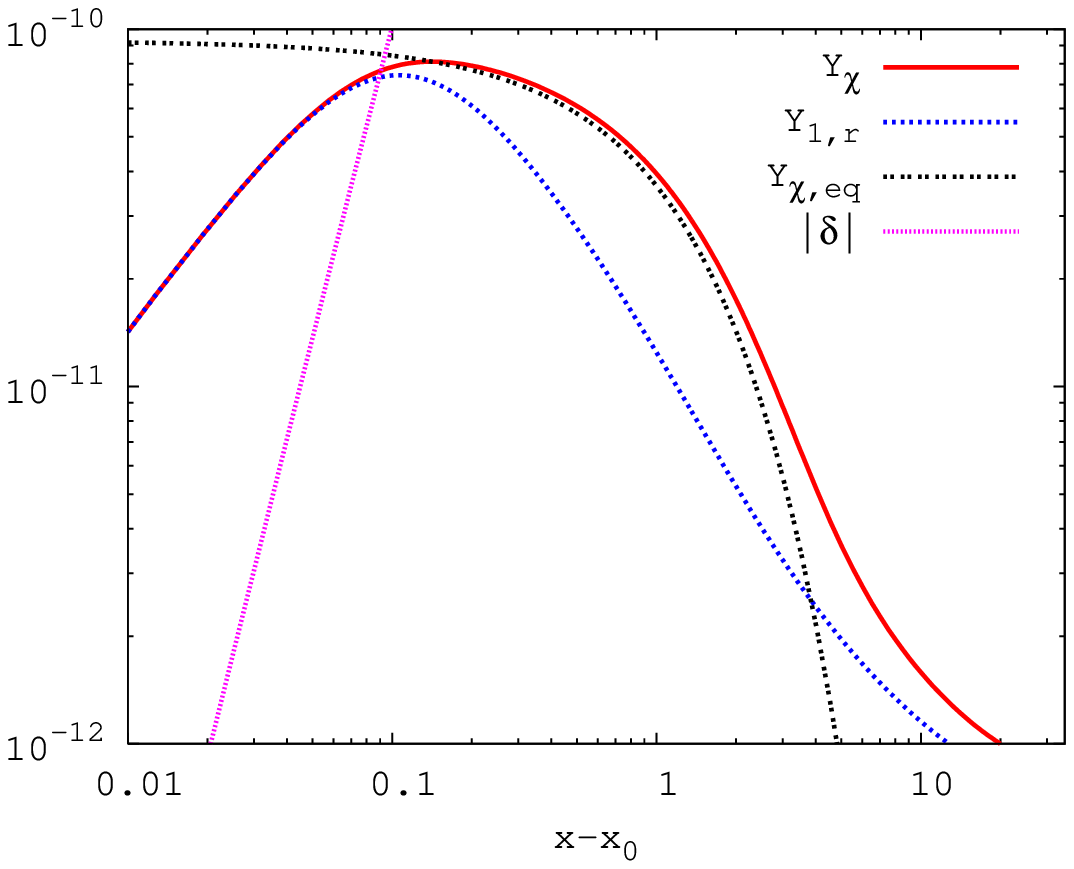}}
    \put(-115,-12){(d)} 
\caption{\footnotesize Evolution of the exact solution $Y_\chi$ (solid red
  curves), $Y_{1,r}$ of Eq.(\ref{eq:resum}) (dotted blue), the equilibrium
  density $Y_{\chi,{\rm eq}}$ of Eq.(\ref{eq:eq}) (double--dotted black), and
  $|\delta|$ of Eq.(\ref{eq:delta1}) (short--dashed violet) as function of
  $x - x_0$. The initial abundance is assumed to be $Y_\chi(x_0=22) = 0$.  We
  take (a) 
  $a=10^{-9}$ GeV$^{-2}$, $b=0$, (b) $a=10^{-8}$ GeV$^{-2}$, $b=0$, (c) $a=0$,
  $b=10^{-8}$ GeV$^{-2}$, and (d) $a=0$, $b=10^{-7}$ GeV$^{-2}$. In frames (a)
  and (c) the curves for $Y_{1,r}$ practically coincide with the solid lines.}
    \label{fig:abundance_x}
  \end{center}
\end{figure}

In Fig.~\ref{fig:abundance_x} we present the evolution of the exact, numerical
solution $Y_\chi$ (solid red), $Y_{1,r}$ (dotted blue), $Y_{\chi,{\rm eq}}$
(double--dotted black) and $|\delta|$ (short--dashed violet) as function of $x
- x_0$. Here we consider vanishing initial $\chi$ density, $Y_\chi(x_0=22) =
0.$ Clearly the first order approximation $Y_1$ of Eq.(\ref{eq:first_order})
fails to reproduce the exact result once $|\delta|$ becomes comparable to
$Y_0$. On the contrary, frames (a) and (c) show that the re--summed ansatz
$Y_{1,r}$ of Eq.(\ref{eq:resum}) reproduces the numerical solution very well
for all $x > x_0$ if $a \lsim 10^{-9}$ GeV$^{-2}$ and $b \lsim 10^{-8}$
GeV$^{-2}$.  However, for intermediate values of $x-x_0$, the disagreement
between $Y_{1,r}$ and the exact solution becomes large as the cross section
increases. In frames (b) and (d) of Fig.~\ref{fig:abundance_x} sizable
deviations from the exact value are observed at $x-x_0 \sim 1$ for $a=
10^{-8}$ GeV$^{-2}$ or $b = 10^{-7}$ GeV$^{-2}$. For larger $x$ the deviation
becomes smaller again, and for $x \gg x_0$ the difference is insignificant
even for these large cross sections.

\begin{figure}[t!]
  \begin{center} 
    \hspace*{-0.5cm}
    \scalebox{0.63}{\includegraphics*{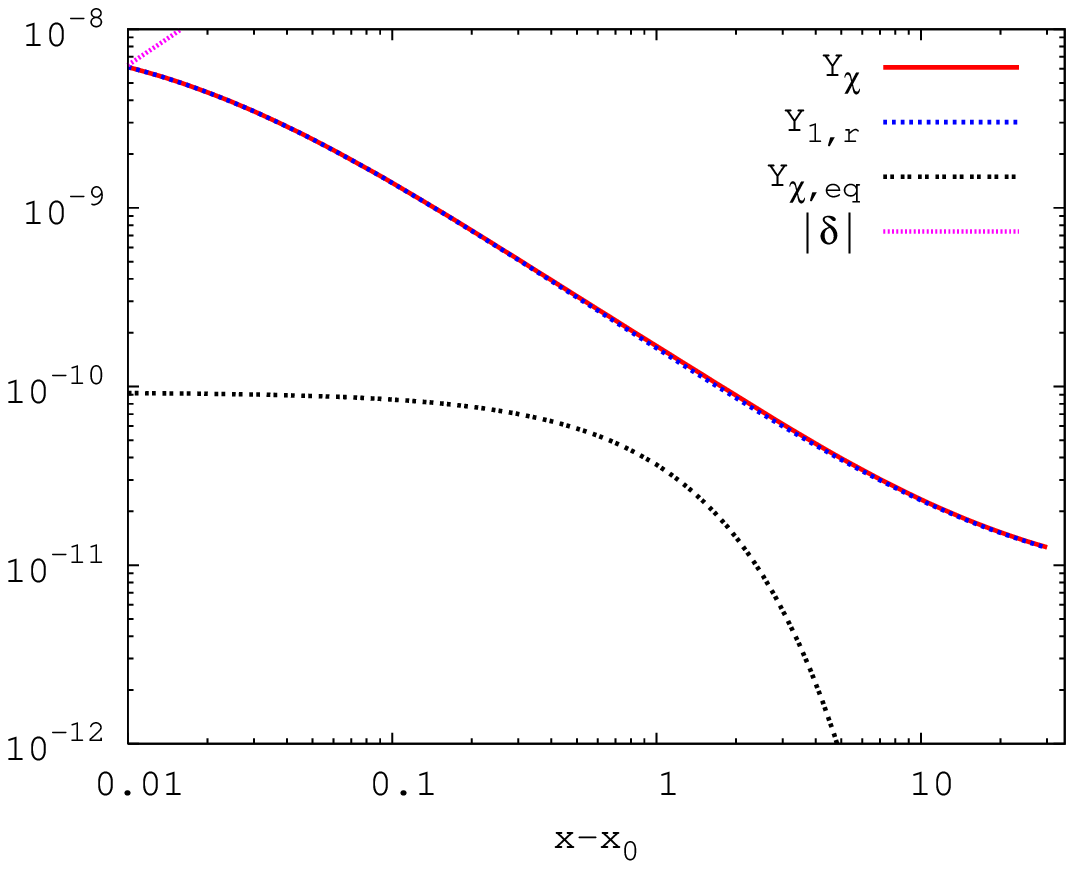}}
    \put(-115,-12){(a)} \hspace*{-0.5cm}
    \scalebox{0.63}{\includegraphics*{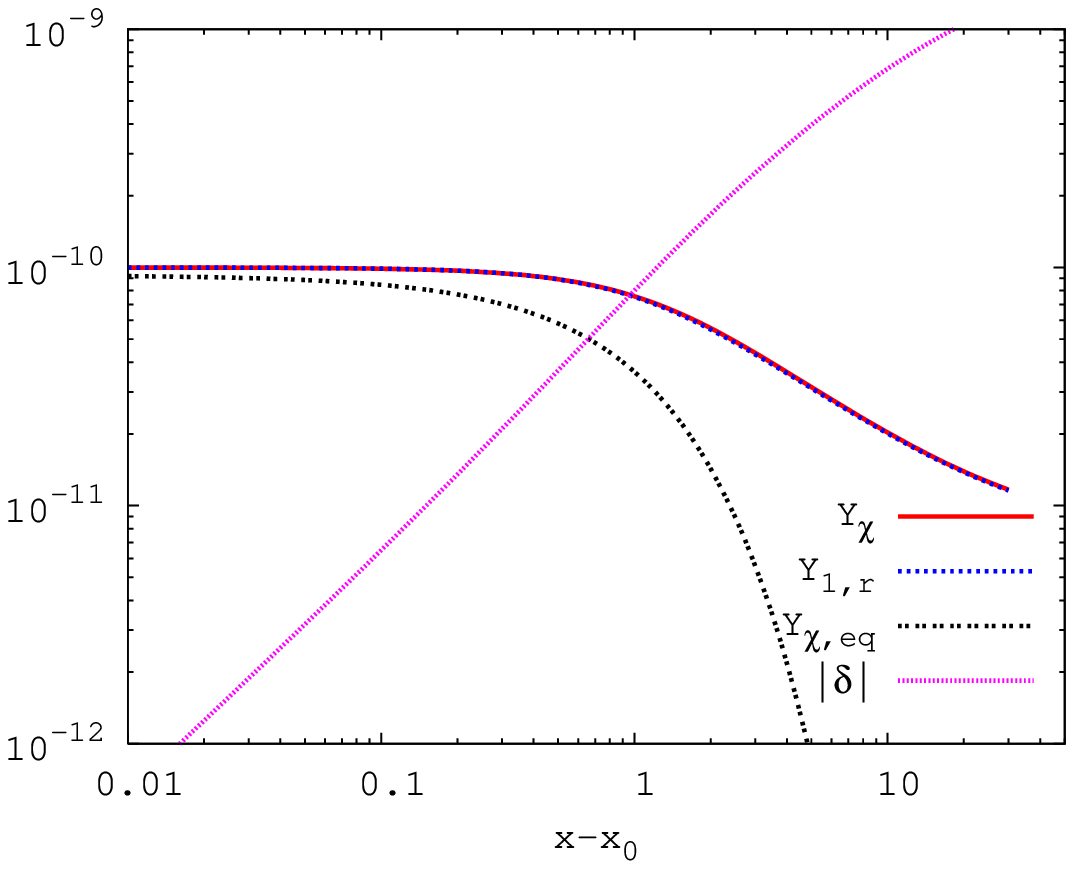}}
    \put(-115,-12){(b)} \vspace{0.5cm} \hspace*{-0.5cm}
    \scalebox{0.63}{\includegraphics*{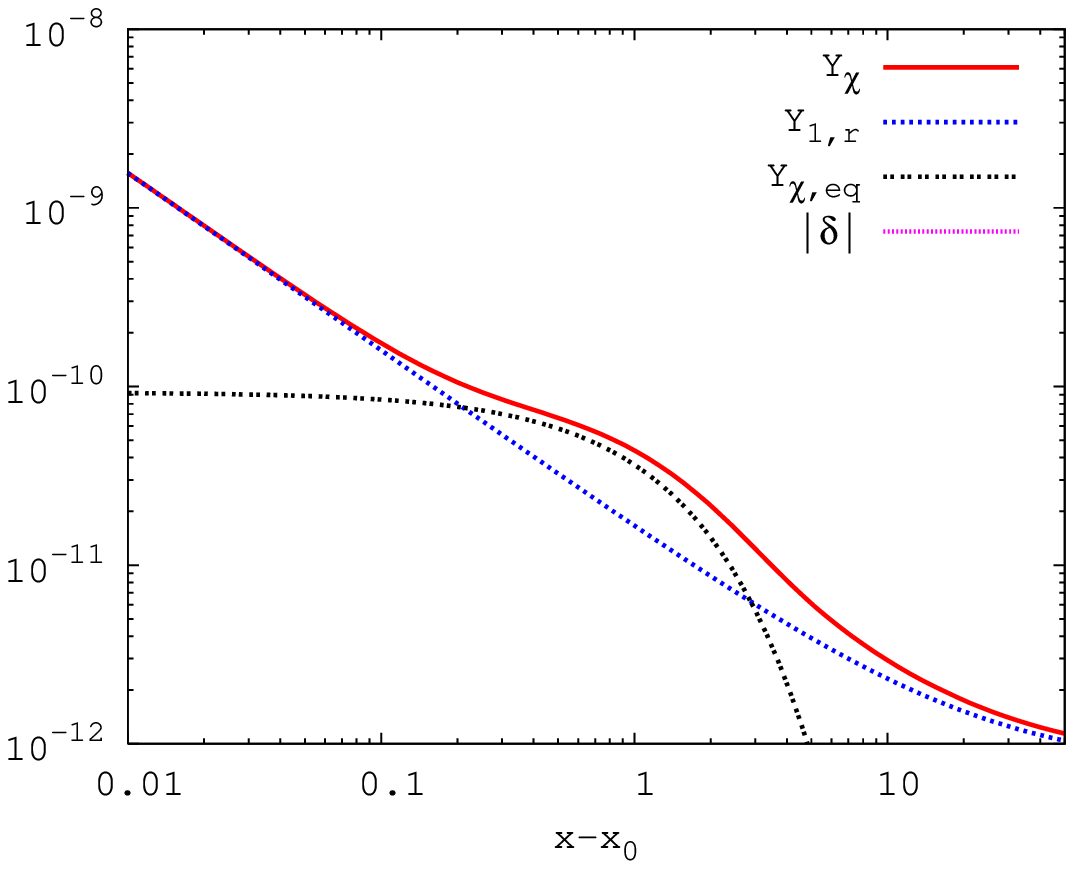}}
    \put(-115,-12){(c)} \hspace*{-0.5cm}
    \scalebox{0.63}{\includegraphics*{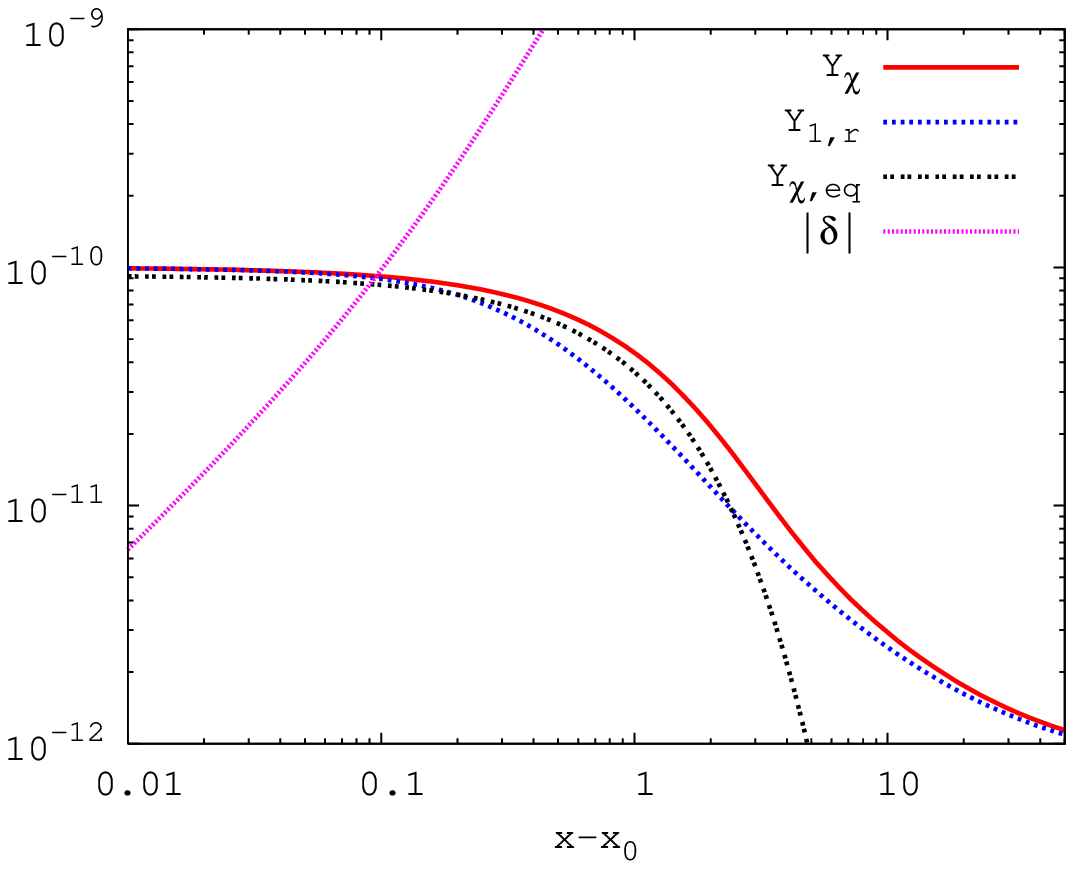}}
    \put(-115,-12){(d)} 
\caption{\footnotesize Evolution of $Y_\chi$ (solid red curves), $Y_{1,r}$
  (dotted blue), $Y_{\chi,{\rm eq}}$ (double--dotted black) and $|\delta|$
  (short--dashed violet) as function of $x - x_0$.
  Here we take (a) $a=10^{-9}$ GeV$^{-2}$, $Y_\chi(x_0)=10^{-8}$, (b)
  $a=10^{-9}$ GeV$^{-2}$, $Y_\chi(x_0)=10^{-10}$, (c) $a=10^{-8}$ GeV$^{-2}$,
  $Y_\chi(x_0)=10^{-7}$ and (d) $a=10^{-8}$ GeV$^{-2}$, $Y_\chi(x_0) =
  10^{-10}$. The other parameters are as in Fig.~\ref{fig:abundance_x}.}
    \label{fig:nonzero}
  \end{center}
\end{figure}

We also analyzed scenarios with sizable initial $\chi$ abundance, $Y_\chi(x_0)
\neq 0$. Figure~\ref{fig:nonzero} shows that the re--summed ansatz again
matches the numerical result very well for all values of $x$ if $a \lsim
10^{-9}$ GeV$^{-2}$. This is not surprising since, as we saw in the discussion
of Eq.(\ref{eq:check}), it reproduces the exact solution if $Y_\chi(x_0)$
dominates over the thermal contribution. For $a = 10^{-8}$ GeV$^{-2}$,
$Y_{1,r}$ again starts to deviate from the exact numerical solution at $x \sim
0.1$, but approaches it for $x \gg x_0$. Note also that already for the
smaller cross section chosen in this Figure, the final relic density is almost
independent of $Y_\chi(x_0)$.

\vspace*{5mm}

\begin{figure}[h!]
  \begin{center}
    \hspace*{-1.5cm}
    \scalebox{0.80}{\includegraphics*{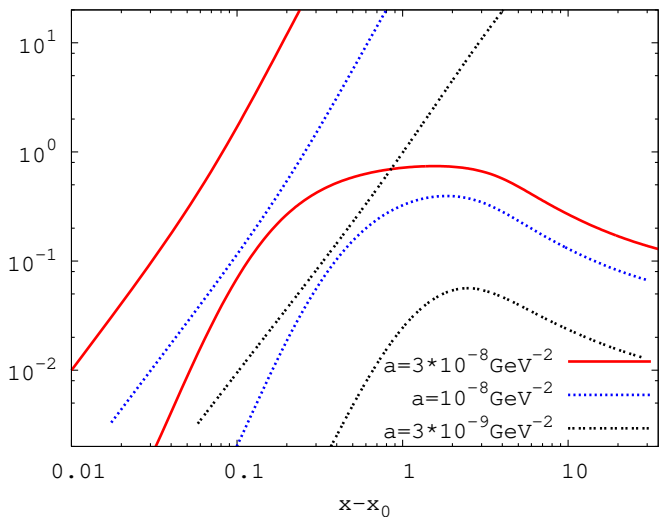}}
\caption{\footnotesize Evolution of $|\delta|/Y_\chi$ (upper curves) and 
  $\epsilon/Y_\chi$ (lower curves) as function of $x - x_0$ for $a=3 \times
  10^{-8}$ GeV$^{-2}$ (solid red), $a=10^{-8}$ GeV$^{-2}$ (dotted blue) and
  $a=3 \times 10^{-9}$ GeV$^{-2}$ (double--dotted black).  Here we
  choose $b=0$ and $Y_\chi(x_0=22)=0$.  }
    \label{fig:epsilon}
  \end{center}
\end{figure}

Let us take a closer look at the difference between the exact solution and the
re--summed ansatz. To this end, we define the deviation $\epsilon$ by
\begin{eqnarray}
  Y_\chi = \frac{Y_0}{1 - \delta/Y_0} + \epsilon\, .
\end{eqnarray}
Inserting this ansatz into the Boltzmann equation (\ref{eq:boltzmann}) leads
to the evolution equation for $\epsilon$:
\begin{eqnarray} \label{eq:epsilon}
  \frac{d\epsilon}{dx} = - \frac {f \langle \sigma v \rangle}{x^2}
  \left[ \epsilon^2 + 2 \epsilon\ \frac{Y_0}{1 - \delta/Y_0}
    - \frac{(\delta/Y_0)^2}{(1 - \delta/Y_0)^2}\ Y_{\chi,{\rm eq}}^2 \right]\,
  ,
\end{eqnarray}
which again resembles the Boltzmann equation. Since initially $\epsilon=0$,
our re--summed ansatz works very well as long as $\delta/Y_0$ remains
suppressed.  Note that the inhomogeneous term on the rhs of
Eq.(\ref{eq:epsilon}) is of order $(\delta/Y_0)^2$. The analogous correction
to our original first order solution $Y_1$ of Eq.(\ref{eq:first_order}) would
start at ${\cal O}(\delta/Y_0)$. Since this inhomogeneous term is positive,
$\epsilon(x) > 0$ for all $x > x_0$, i.e. $Y_{1,r}$, like $Y_1$, always
under--estimates the exact solution.  As $|\delta|/Y_0$ grows, the last term
in Eq.(\ref{eq:epsilon}) can become sizable. Note, however, that it is
multiplied with $\left( Y_{\chi,{\rm eq}} \right)^2$, which drops $\propto
\exp{(-2x)}$ with increasing $x$.  Therefore $\epsilon$ becomes large only if
$|\delta|$ reaches values of order of $Y_0$ for $x-x_0 \lsim 1$. The
homogeneous 
terms in Eq.(\ref{eq:epsilon}) imply that for large $x - x_0$ the deviation
$\epsilon$ decreases again, similar to the WIMP relic abundance $Y_{\chi}$.
This situation is depicted in Fig.~\ref{fig:epsilon}, which shows the
evolutions of $|\delta|/Y_\chi$ (upper curves) and $\epsilon/Y_\chi$ (lower
curves) as function of $x - x_0$ for $a=3 \times 10^{-8}$ GeV$^{-2}$ (solid
red), $a=10^{-8}$ GeV$^{-2}$ (dotted blue) and $a=3 \times 10^{-9}$ GeV$^{-2}$
(double--dotted black).  Here we choose $b=0$ and $Y_\chi(x_0=22)=0$.
Even in the case where $\epsilon$ becomes sizable for intermediate values of
$x$, it eventually diminishes and hence our analytical formula succeeds in
reproducing the present relic abundance $Y_\chi(x \rightarrow \infty)$ fairly
well.

\begin{figure}[h!]
  \begin{center}
    \hspace*{-0.5cm} 
    \scalebox{0.62}{\includegraphics*{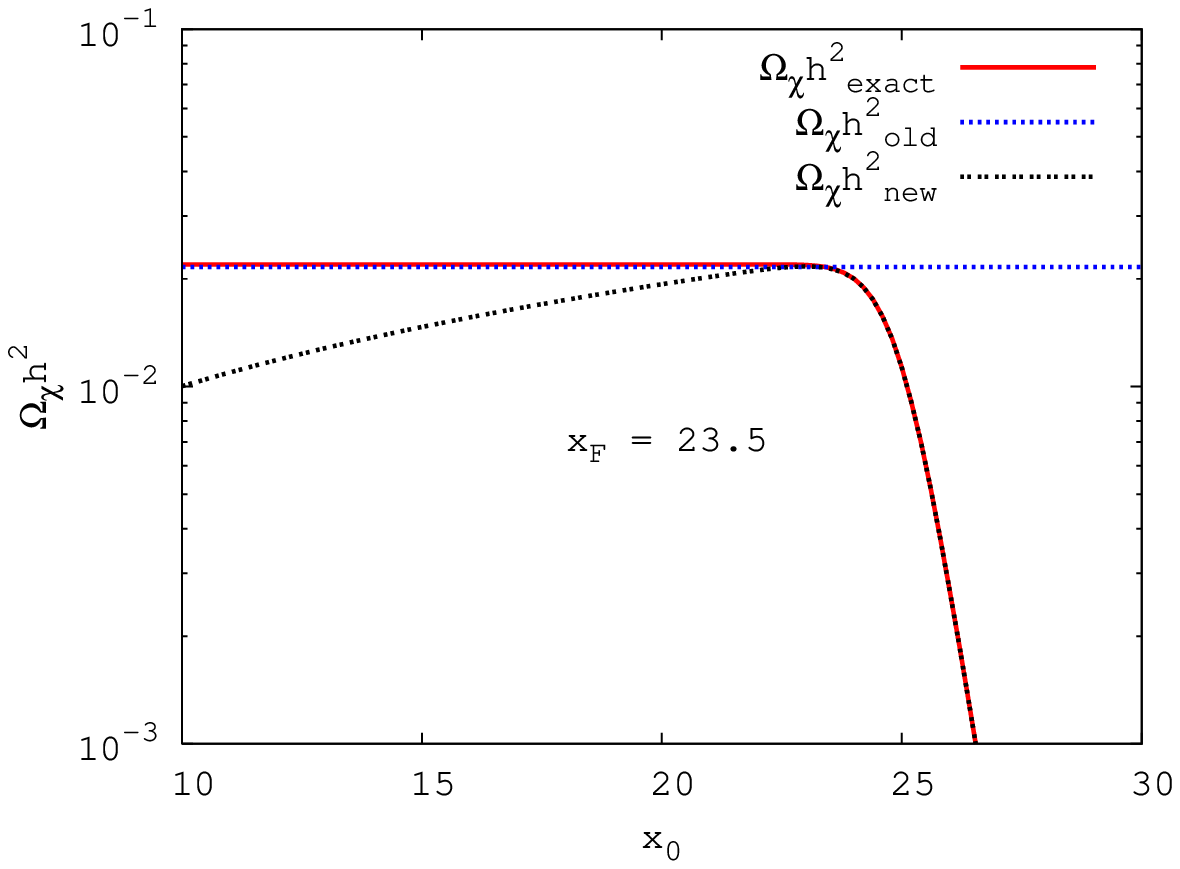}}
    \put(-115,-12){(a)} \hspace*{-0.5cm}
    \scalebox{0.62}{\includegraphics*{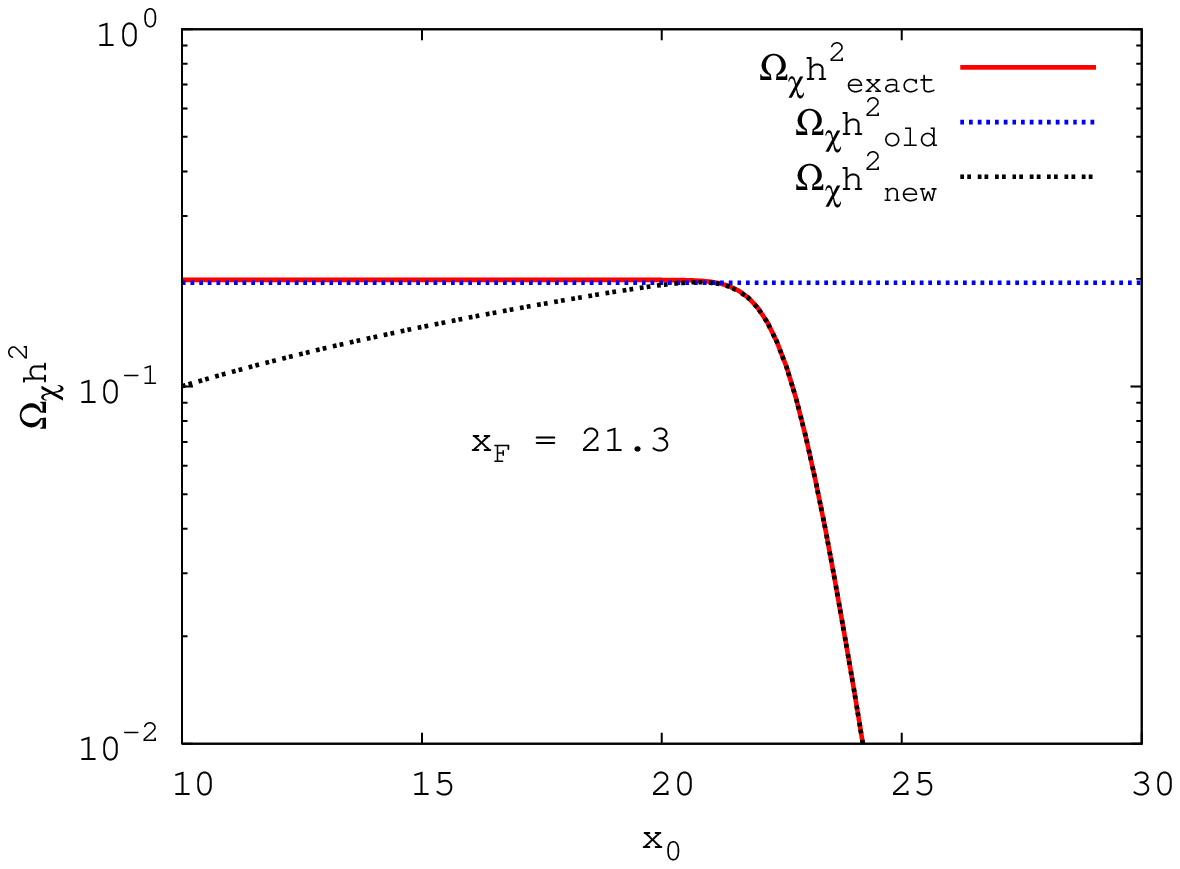}}
    \put(-115,-12){(b)} \hspace*{-0.5cm}
\caption{\footnotesize The present relic density evaluated numerically (solid
  red curves), the old standard approximation (dotted blue) and our new
  approximation (double--dotted black) as function
  of $x_0$. Here we take (a) $a=10^{-8}$ GeV$^{-2}$, $b=0$ and (b) $a=10^{-9}$
  GeV$^{-2}$, $b=0$.}
    \label{fig:comparison}
  \end{center}
\end{figure}

Let us turn to a discussion of the dependence of the present relic abundance
on the initial temperature. In Fig.~\ref{fig:comparison} we plot the present
relic density evaluated numerically (solid red curves), the old standard
approximation (dotted blue) and our new approximation (double--dotted black)
as function of $x_0$. Here we take (a) $a=10^{-8}$ GeV$^{-2}$, $b=0$ and (b)
$a=10^{-9}$ GeV$^{-2}$, $b=0$. We find that our approximation agrees with the
exact result very well for $x_0> x_F$. On the other hand, for $x_0 < x_F$, our
approximation gives too small an abundance\footnote{For $x_0 \ll x_F$, our
  expressions predict $\Omega_\chi h^2 \propto x_0$.} while the old
approximation works very well. The transition between the two regimes is very
sharp. For $x_0 = x_F + 2$, the old approximation over--estimates the relic
abundance by as much as an order of magnitude, while for $x_0 = x_F$ both the
old and the new approximation work well.

\begin{figure}[h!]
  \begin{center}
    \hspace*{-0.5cm} 
    \scalebox{0.62}{\includegraphics*{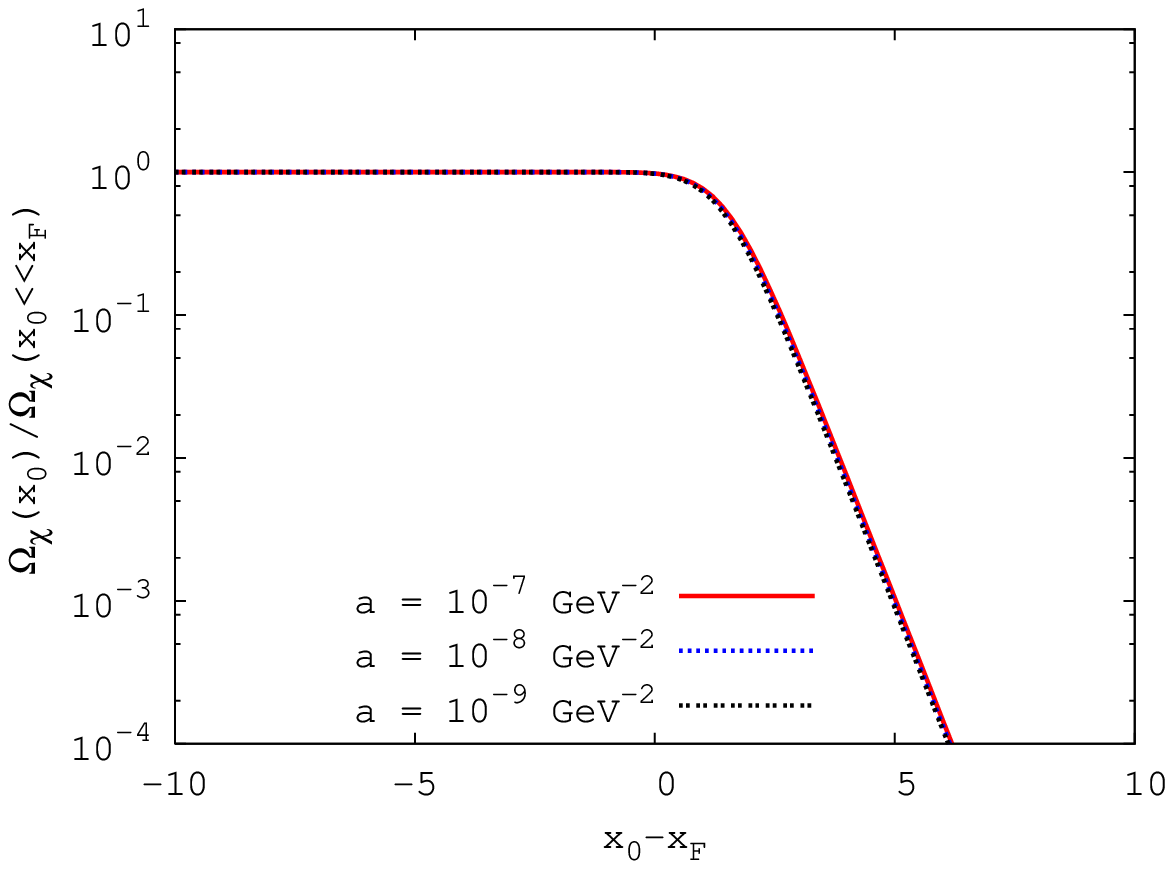}}
    \put(-115,-12){(a)} \hspace*{-0.5cm}
    \scalebox{0.62}{\includegraphics*{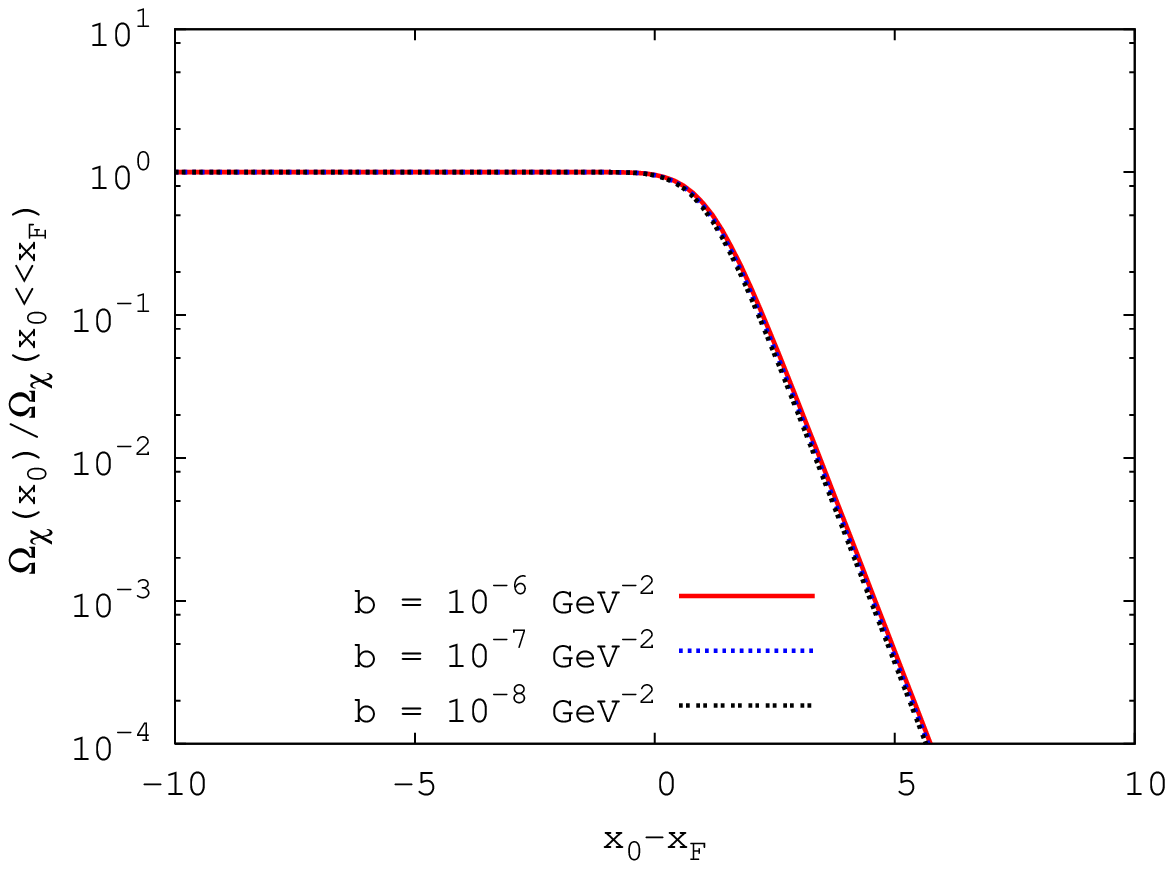}}
    \put(-115,-12){(b)} \hspace*{-0.5cm}
\caption{\footnotesize $\Omega_\chi(x_0)/\Omega_\chi(x_0 \ll x_F)$ as
  function of $x_0-x_F$. In the left frame, $a=10^{-7}$ GeV$^{-2}$ (solid red
  curves), $10^{-8}$ GeV$^{-2}$ (dotted blue) and $10^{-9}$ GeV$^{-2}$
  (double--dotted black) with $b=0$, whereas in the right frame,
  $b=10^{-6}$ GeV$^{-2}$ (solid red), $10^{-7}$ GeV$^{-2}$ (dotted green) and 
  $10^{-8}$ GeV$^{-2}$ (double--dotted black) with $a=0$.}
    \label{fig:scaling}
  \end{center}
\end{figure}

We found that for vanishing initial $\chi$ density, $Y_\chi(x_0) = 0$,
different values of the cross section lead to a universal behavior when the
present relic density is expressed as function of $x_0 - x_F$ and in units
of the relic density for $x_0 \ll x_F$. This can be seen from the analytic
solution we have obtained.  For $x_0 \ll x_F$ it is obvious that
$\Omega_\chi(x_0)/\Omega_\chi(x_0 \ll x_F)$ is nothing but unity and
independent of the cross section. For $x_0 \gg x_F$, the exact solution is
roughly given by the zeroth order approximation $Y_0$, which scales like $a
{\rm e}^{-2x_0}$ if $a$ dominates and the initial abundance vanishes.
Meanwhile, Eq.(\ref{eq:x_F}) shows that $x_F$ is roughly proportional to $\ln
a$. Therefore we obtain the relation
\begin{eqnarray}
  \frac{\Omega_\chi(x_0)}{\Omega_\chi(x_0 \ll x_F)}
  \propto \frac{a {\rm e}^{-2x_0}}{1/a} \propto {\rm e}^{-2(x_0 - x_F)}\, ,
\end{eqnarray}
which has no explicit dependence on the cross section. The same argument is
applicable to the case where $b$ is dominant.  In Fig.~\ref{fig:scaling} we
plot the ratio of the exact present relic density to the value for $x_0 \ll
x_F$, $\Omega_\chi(x_0)/\Omega_\chi(x_0 \ll x_F)$, as function of $x_0 -
x_F$ for various values of $a$ and $b$.  These figures clearly show the
expected scaling behavior both for $a \neq 0, \, b = 0$ (left frame) and for
$a = 0, \, b \neq 0$ (right frame). However, for $Y_\chi(x_0) \neq 0$, no such
scaling exists, apart from the fairly obvious result that $Y_\chi(x \gg x_0)$
becomes independent of $Y_\chi(x_0)$ if $x_0 \ll x_F$.

Fig.~\ref{fig:comparison} shows that $Y_{1,r}(x_0, x\rightarrow \infty)$ has a
well defined maximum when $x_0$ is varied. This maximum occurs at a value
$x_{0,{\rm max}}$ which is close, but not identical, to the decoupling
temperature $x_F$ of Eq.(\ref{eq:x_F}). From the asymptotic expressions for
$Y_0$, Eq.(\ref{eq:y0as}), and $\delta$, Eq.(\ref{eq:deltaas}), we find for
$Y_\chi(x_0) = 0$:
\begin{eqnarray} \label{eq:xmax}
 x_{0, {\rm max}} & \simeq & 
 \frac{1}{2} \ln \frac{f^2 c (a + 6b/x_{0,{\rm max}})^2}{4 x_{0,{\rm max}} }
\nonumber \\
& = & \ln \frac{0.096\ m_\chi M_{\rm Pl} g_\chi(a +
  6b/x_{0,{\rm max}})}{\sqrt{x_{0,{\rm max}} g_*}}\, .
\end{eqnarray}
In deriving this equation, we neglect non--leading terms in $1/x_{0,{\rm
    max}}$ in each combination of $a$ and $b$.\footnote{The next--to--leading
  correction to the pure $a-$term would have been relevant, but it cancels.
  The non--leading corrections to terms that require both $a$ and $b$ to be
  non--zero are numerically insignificant, and of the same order as terms
  omitted in the expansion (\ref{eq:ab}) of the annihilation cross section.}
Notice that $x_{0,{\rm max}}$ coincides with $x_F$ of Eq.(\ref{eq:x_F}), if
one chooses $\xi = 1/4$ (rather than $\xi = \sqrt{2}-1$).

Since the actual relic density is already practically independent of $x_0$ for
$x_0 < x_{0,{\rm max}}$ we can construct a new semi-analytic solution which
describes the relic density for the whole range of $x_0$: for $x_0 > x_{0,{\rm
    max}}$, compute the relic density from $Y_{1,r}(x_0)$, but for $x_0 <
x_{0,{\rm max}}$, use $Y_{1,r}(x_{0,{\rm max}})$ instead. 

\begin{figure}[t!]
  \begin{center}
    \hspace*{-0.5cm} 
    \scalebox{0.6}{\includegraphics*{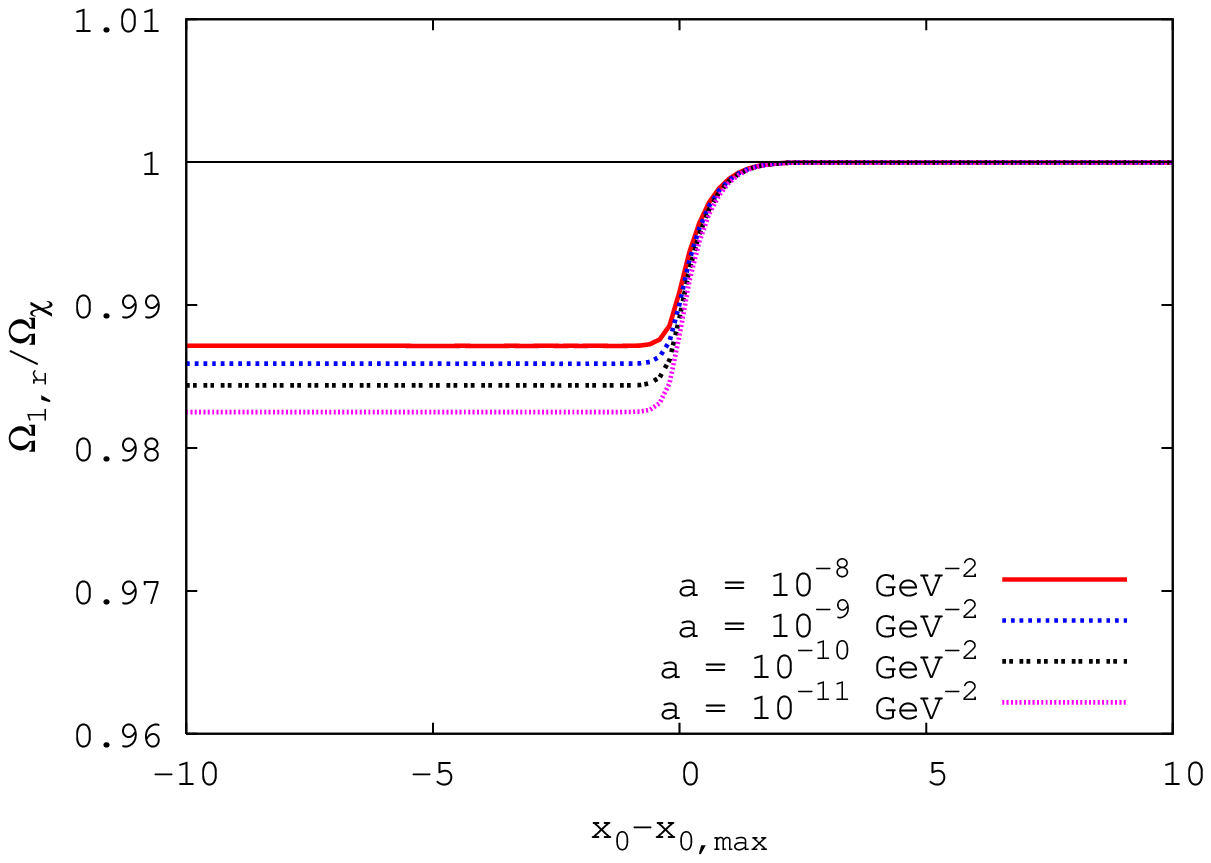}}
    \put(-115,-12){(a)} \hspace*{-0.5cm}
    \scalebox{0.6}{\includegraphics*{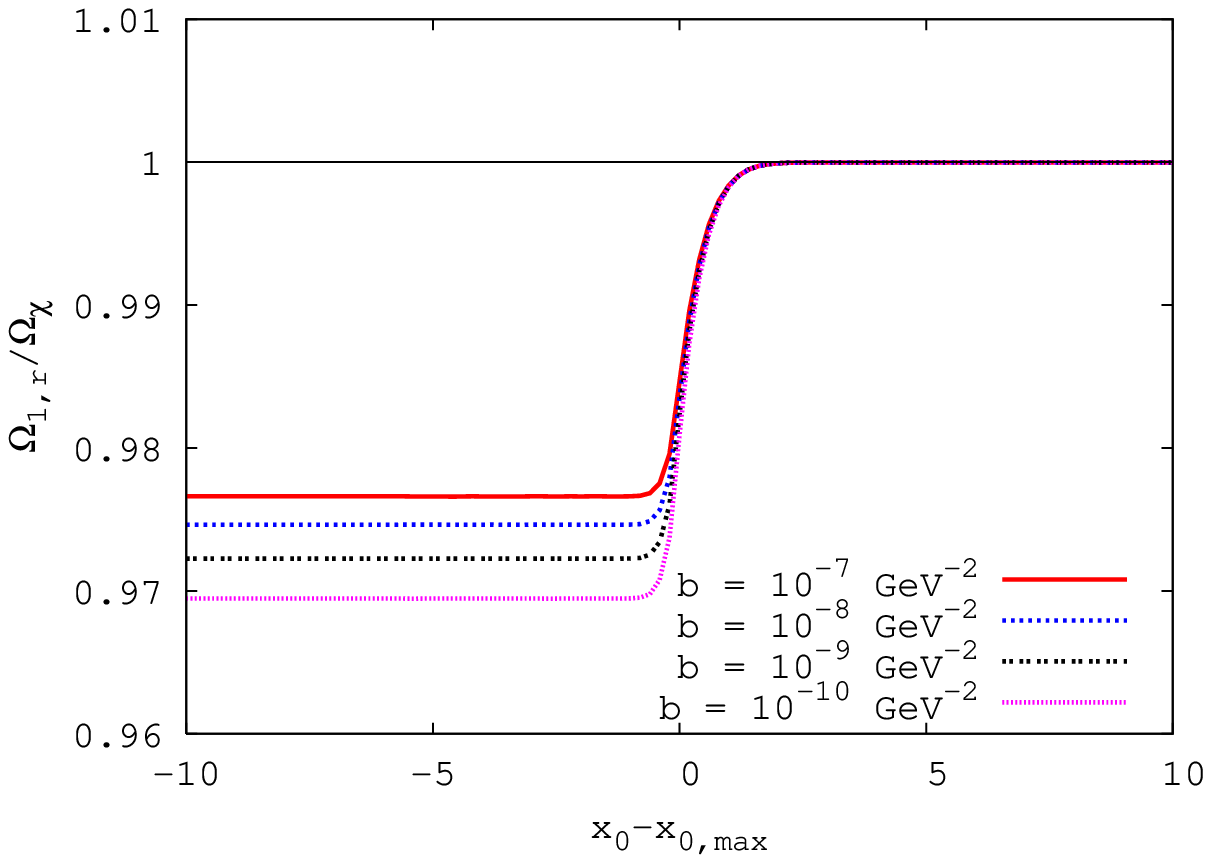}}
    \put(-115,-12){(b)} 
\caption{\footnotesize Ratios of approximate and exact results for the relic 
  density $\Omega_{1,r}/\Omega_\chi$ as function of $x_0 - x_{0,{\rm max}}$,
  for $a \neq 0, \, b = 0$ (left frame) and $a = 0, b \neq 0$ (right frame).
  The curves use $Y_{1,r}$ with $x_0$ replaced by $\max(x_0, \, x_{0,{\rm
      max}})$, see Eq.(\ref{eq:xmax}). In the left frame, $a=10^{-8}$
  GeV$^{-2}$ (solid red curves), $10^{-9}$ GeV$^{-2}$ (dotted blue),
  $10^{-10}$ GeV$^{-2}$ (double--dotted black), $10^{-11}$ GeV$^{-2}$
  (short--dashed violet) with $b=0$, whereas in the right frame, $b=10^{-7}$
  GeV$^{-2}$ (solid red), $10^{-8}$ GeV$^{-2}$ (dotted blue), $10^{-9}$
  GeV$^{-2}$ (double--dotted black), $10^{-10}$ GeV$^{-2}$ (short--dashed
  violet) with $a=0$.
}
    \label{fig:deviation}
  \end{center}
\end{figure}

The ratio of this semi--analytic result $\Omega_{1,r}$ to the exact value
$\Omega_\chi$ is depicted in Fig.~\ref{fig:deviation}. As noted earlier, our
approximation becomes exact for $x_0 \gsim x_F$. For smaller $x_0$ the new
approximation still slightly under--estimates the correct answer, but the
deviation is at most $1.7$\% for $b=0$ (left frame), and $3.0$\% for $a=0$
(right frame). On the other hand, in the same region the old standard
approximation reproduces the present relic abundance within 1\% error.
We thus see that for $x_0 < x_F$, this new expression works nearly as well as
the old 
standard result;\footnote{However, if $a=0$, we should expect ${\cal O}(10\%)$
  corrections to the relic density from higher order terms in the expansion
  (\ref{eq:ab}) of the cross section; if $a \neq 0$, these higher order terms
  should only contribute ${\cal O}(1\%)$.} of course, the old result fails
badly for $x_0 > x_F$.  Finally, since by definition $Y_{1,r}$ depends only
weakly on $x_0$ for $x_0 \sim x_{0,{\rm max}}$, the latter quantity need not
be calculated very precisely; in practice, setting $x_{0,{\rm max}} = 20$ in
the rhs of Eq.(\ref{eq:xmax}) is often sufficient. In contrast, the standard
approximation (\ref{eq:omold}) depends linearly (for $b=0$) or even
quadratically (for $a=0$) on $x_F$; several iterations are therefore required
to solve Eq.(\ref{eq:x_F}) to sufficient accuracy.  Altogether, our new
semi--analytic formula is evidently a quite powerful tool in calculating the
density of cold relics.

\vspace*{5mm}

\section{Relic Abundance Including the Decay of Heavier Particles}

In this section we investigate a scenario where unstable heavy particles
$\phi$ decay into long--lived or stable particles $\chi$. We assume that
$\phi$ decays out of thermal equilibrium, so that $\phi$ production is
negligible; however, we include both thermal and non--thermal production of
$\chi$ particles. For example in some supersymmetric models neutralinos,
which are stable due to R--parity, can be produced non--thermally through the
decay of moduli \cite{moroi} or gravitinos after the end of inflation. The
number 
densities of $\chi$ and $\phi$ obey the following coupled Boltzmann equations:
\begin{eqnarray}
  \frac{dn_\chi}{dt} + 3 H n_\chi & = & - \langle \sigma v \rangle
  (n_\chi^2 - n_{\chi,{\rm eq}}^2 ) + N \Gamma_\phi n_\phi\, , \nonumber \\
  \frac{dn_\phi}{dt} + 3 H n_\phi & = & - \Gamma_\phi n_\phi\, ,
\end{eqnarray}
where $N$ is the average number of $\chi$ particles produced in a $\phi$
decay, and $\Gamma_\phi$ and $n_\phi$ are the decay rate and the number
density of the heavier particle. In contrast to refs.\cite{low_others} we
assume that $\phi$ does {\em not} dominate the total energy density, so that
the co--moving entropy density remains approximately constant throughout. The
Boltzmann equation for $n_\phi$ can then easily be solved analytically, using
the fact that $t \propto T^{-2} \propto x^2$ in the radiation--dominated
era. Inserting this solution into the equation for $n_\chi$, and again
switching variables to $Y_\chi=n_\chi/s$, $Y_\phi=n_\phi/s$ and $x$, the
Boltzmann equation for $\chi$ becomes
\begin{eqnarray}
  \frac{dY_\chi}{dx} = - \frac{\langle \sigma v \rangle s}{H x}
  (Y_\chi^2 - Y_{\chi,{\rm eq}}^2) + N r x Y_\phi(x_0)
  \exp \left( - \frac{r}{2} (x^2 - x_0^2) \right)\, ,
  \label{eq:boltzmann_d}
\end{eqnarray}
where $r = \Gamma_\phi/ H x^2 = (\Gamma_\phi M_{\rm Pl} / \pi m_\chi^2)
\sqrt{90/g_*}$ is constant. The zeroth order solution of
Eq.(\ref{eq:boltzmann_d}) is again obtained by neglecting $\chi$
annihilation. Using the expansion (\ref{eq:ab}) of the annihilation cross
section, we have
\begin{eqnarray}
  \frac{d Y_0}{dx} = f \left(a + \frac{6b}{x} \right)  c x {\rm e}^{-2x}
  + N r x Y_\phi(x_0)
  \exp \left( - \frac{r}{2} (x^2 - x_0^2) \right)\, .
\end{eqnarray}
This equation can be integrated, giving
\begin{eqnarray} \label{eq:y0d}
  Y_0 & = & f c \left[ \frac{a}{2} (x_0 {\rm e}^{-2 x_0} - x {\rm e}^{-2 x} )
    + \left( \frac{a}{4} + 3 b \right) ({\rm e}^{-2 x_0} - {\rm e}^{-2 x}) \right]
  \nonumber \\
  &+& N Y_\phi(x_0)
  \left[ 1- \exp \left( - \frac{r}{2}(x^2-x_0^2) \right)
  \right] + Y_\chi(x_0)\, .
\end{eqnarray}
For $x \gg x_0$, $Y_0$ becomes constant,
\begin{eqnarray}
  Y_{0,\infty} = f c \left[ \frac{a}{2} x_0 {\rm e}^{-2 x_0}
    + \left( \frac{a}{4} + 3 b \right) {\rm e}^{-2 x_0} \right]
  + N Y_\phi(x_0) + Y_\chi(x_0)\, .
\end{eqnarray}

For sufficiently large $Y_0$ the annihilation term in
Eq.(\ref{eq:boltzmann_d}) becomes significant. We add a correction term to
include this effect, as in Eq.(\ref{eq:first_order}). Since the new,
non--thermal contribution to $\chi$ production is already fully included in
$Y_0$, the Boltzmann equation for $\delta$ is again given by
Eq.(\ref{eq:eq_delta}). Using now Eq.(\ref{eq:y0d}) for $Y_0$, we can
integrate Eq.(\ref{eq:eq_delta}), giving
\begin{eqnarray} \label{eq:delta_d}
\delta & = & \left\{ - f^3 c^2 \left[ \frac{1}{4} a^3 F^4_0(x,x_0)
  + \frac{1}{4} a^2(a+18b) F^4_1(x,x_0) \right. \right.
  \nonumber \\
  && \hspace*{14mm} \left. + \frac{1}{16} a (a + 12 b)(a + 36 b) F^4_2(x,x_0)
  + \frac{3}{8} b (a + 12 b)^2 F^4_3(x,x_0) \right] \nonumber \\
  && + Y_{0,\infty} f^2 c \left[ a^2 F^2_1(x,x_0)
    + \frac{1}{2} a (a + 24 b) F^2_2(x,x_0)
+ 3 b (a + 12 b) F^2_3(x,x_0) \right] \nonumber \\
  && \left. - Y_{0,\infty}^2 f \left[ a F^0_2(x,x_0)
      + 6b F^0_3(x,x_0) \right] \right. {\bigg\}} \nonumber \\
  &-& N^2 Y^2_\phi(x_0) {\rm e}^{r x_0^2} f [a G_2^r (x,x_0) + 6b G_3^r (x,x_0) ]
  \nonumber \\
  &+& 2 N Y_\phi(x_0) {\rm e}^{r x_0^2/2} Y_{0,\infty} f
  [a G_2^{r/2}(x,x_0) + 6b G_3^{r/2}(x,x_0)]
  \\
  &-& N Y_\phi(x_0) {\rm e}^{r x_0^2/2} f^2 c \left[ a^2 G_1^c(x,x_0)
  + \frac{1}{2} a (a + 24 b) G_2^c(x,x_0)
+ 3 b (a + 12 b) G_3^c(x,x_0) \right]\, . \nonumber
\end{eqnarray}
The functions $G_n^r(x,x_0)$, $G_n^{r/2}(x,x_0)$ and $G_n^c(x,x_0)$ are
defined by 
\begin{eqnarray}
G_n^r(x,x_0) &=& \int_{x_0}^x dt \ \frac{{\rm e}^{- r t^2}}{t^n}\, ,
  \quad n=2,3\, , \nonumber \\
G_n^{r/2}(x,x_0) &=& \int_{x_0}^x dt \ \frac{{\rm e}^{- r t^2/2}}{t^n}\, ,
  \quad n=2,3\, , \nonumber \\
G_n^c(x,x_0) &=& \int_{x_0}^x dt \ \frac{{\rm e}^{-2t - r t^2/2}}{t^n}\, ,
  \quad n=1,2,3\, .
\end{eqnarray}
Explicit expressions for these functions are given in the Appendix,
Eqs.(\ref{eq:gi}). Notice that the expression in curly brackets $\{...\}$ in
Eq.(\ref{eq:delta_d}) has the same form as in Eq.(\ref{eq:delta1}).

\begin{figure}[t!]
  \begin{center}
    \hspace*{-0.5cm}
    \scalebox{0.63}{\includegraphics*{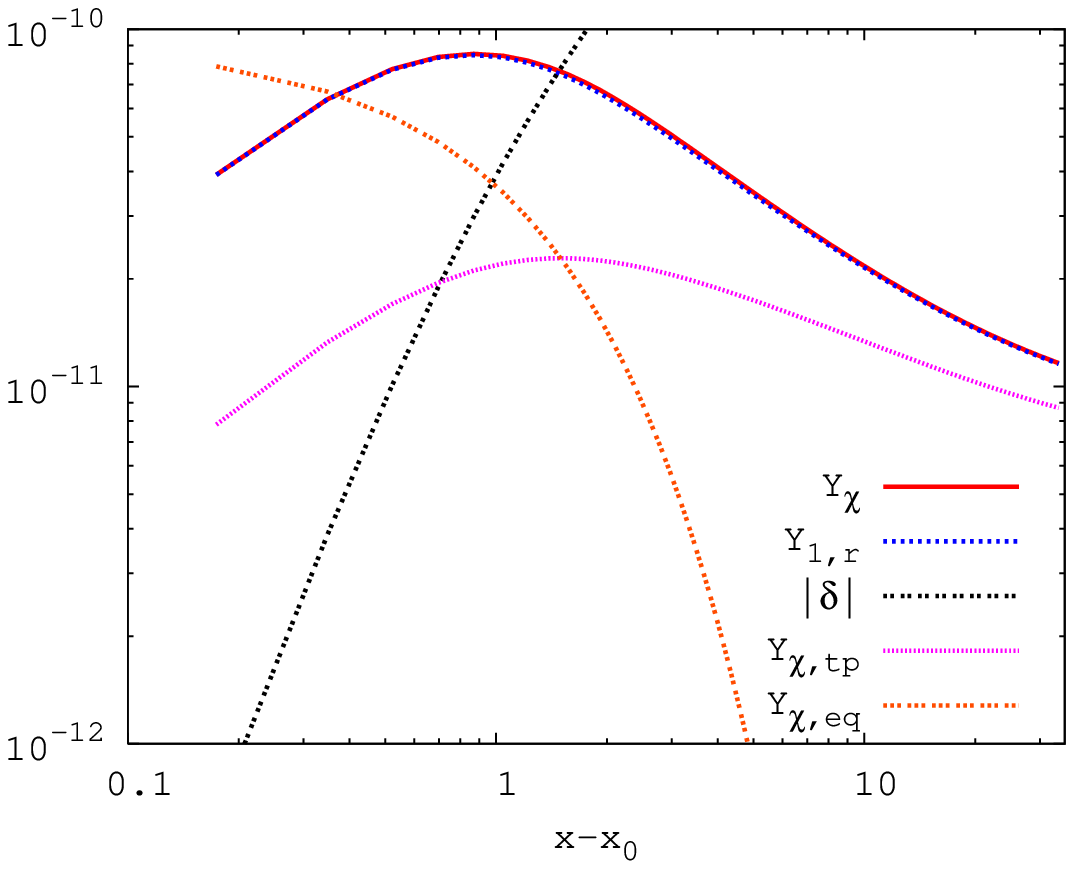}}
    \put(-115,-12){(a)}
    \hspace*{-0.5cm}
    \scalebox{0.63}{\includegraphics*{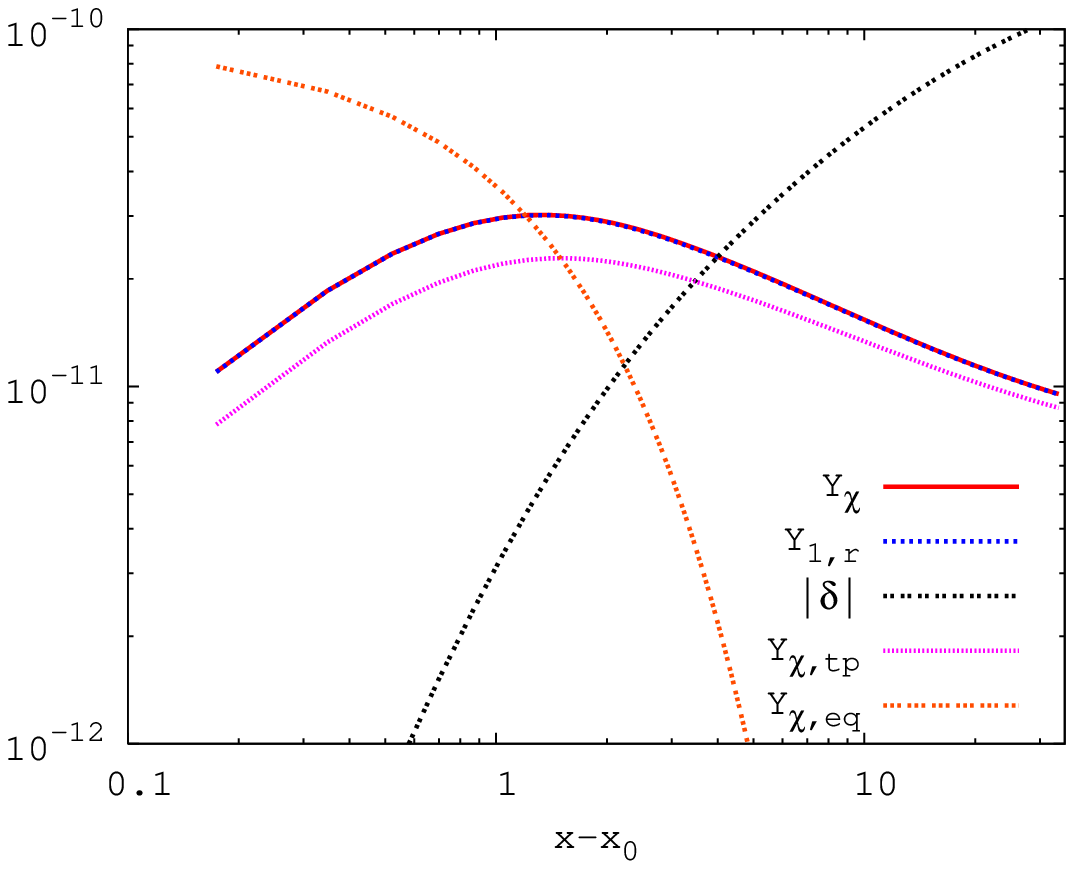}}
    \put(-115,-12){(b)}
    \vspace{0.5cm}
    \hspace*{-0.5cm}
    \scalebox{0.63}{\includegraphics*{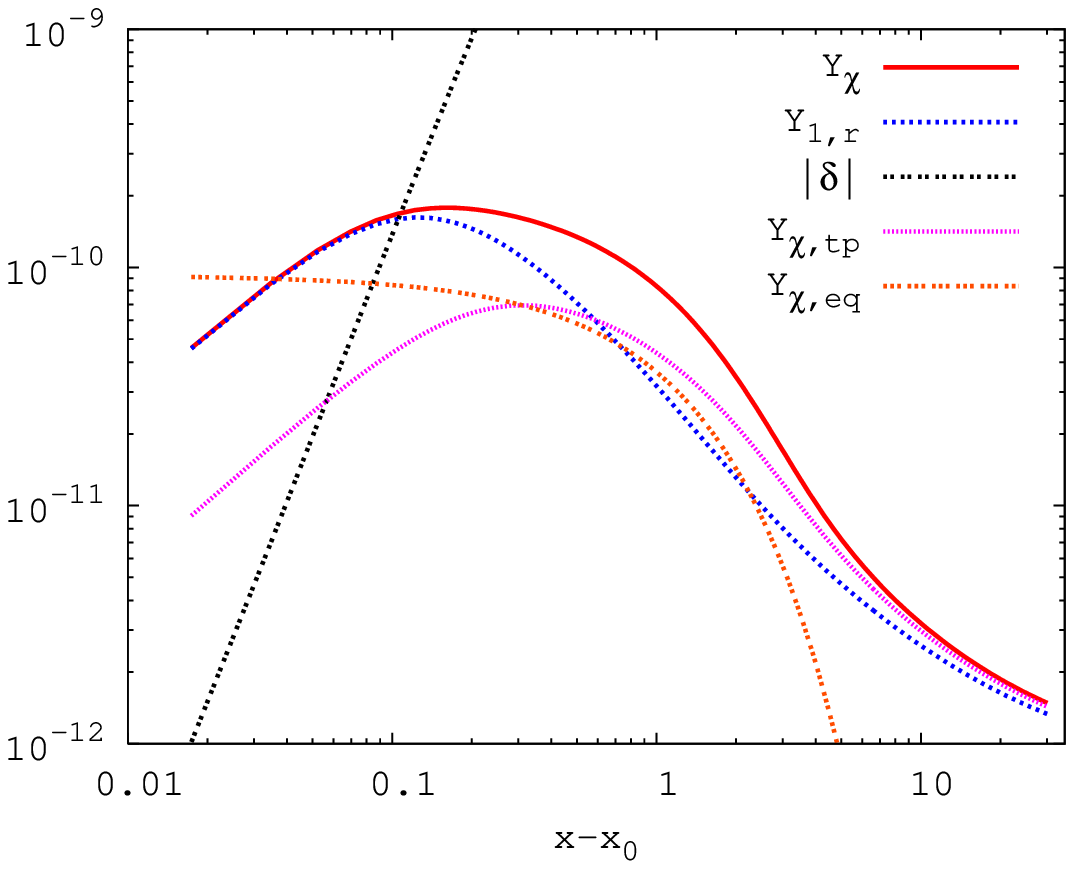}}
    \put(-115,-12){(c)}
    \hspace*{-0.5cm}
    \scalebox{0.63}{\includegraphics*{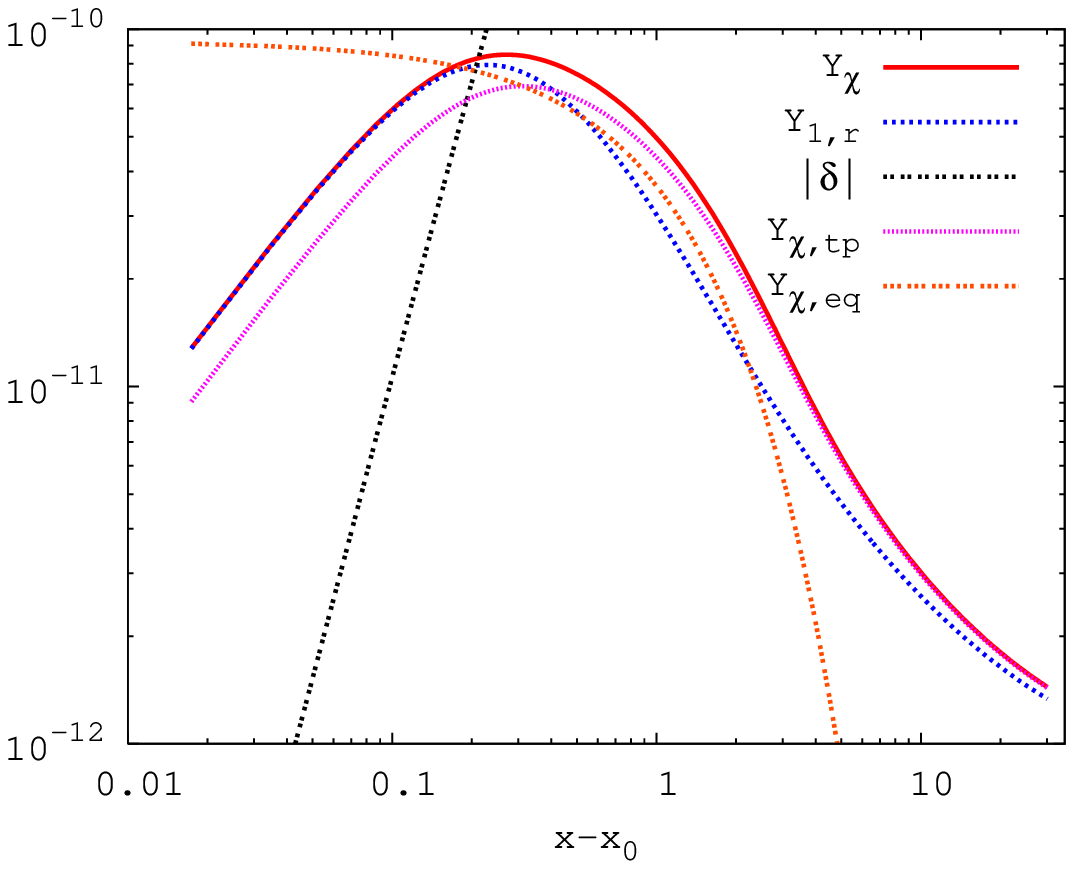}}
    \put(-115,-12){(d)}
\caption{\footnotesize Evolution of $Y_\chi$ (solid red curves), 
  $Y_{1,r}$ (dotted blue), $|\delta|$ (double--dotted black), the prediction
  for purely thermal $\chi$ production $Y_{\chi,{\rm tp}}$ (short--dashed
  violet) and $Y_{\rm eq}$ (triple--dotted orange) as function of $x - x_0$,
  for $Y_\chi(x_0=22)=0$, $r=0.1$, $N=1$ and $b=0$.
  The $S-$wave cross section and the initial $\phi$ density are (a)
  $a=10^{-9}$ GeV$^{-2}$, $Y_\phi(x_0)=10^{-10}$, (b) $a=10^{-9}$ GeV$^{-2}$,
  $Y_\phi(x_0)=10^{-11}$, (c) $a=10^{-8}$ GeV$^{-2}$, $Y_\phi(x_0)=10^{-9}$
  and (d) $a=10^{-8}$ GeV$^{-2}$, $Y_\phi(x_0)=10^{-10}$.}
\label{fig:abundance_d}
\end{center}
\end{figure}

Results for this scenario with $b=0$ are shown in Fig.~\ref{fig:abundance_d}.
We choose $r = 0.1$ so that $r x_0^2 \sim x_0$, which leads to the most
difficult situation where thermal and non--thermal production occur
simultaneously. We see that even for the smaller cross section considered, $a
= 10^{-9}$ GeV$^{-2}$ (top frames), the simple first--order solution
(\ref{eq:first_order}) soon fails, since $|\delta|$ exceeds $Y_0$. However,
the re--summed ansatz $Y_{1,r}$ of Eq.(\ref{eq:resum}) describes the exact
temperature dependence very well for this cross section, both for large (top
left frame) and moderate (top right) non--thermal $\chi$ production. For $a =
10^{-8}$ GeV$^{-2}$ (bottom frames) we again observe sizable deviations for
intermediate values of $x-x_0$. 

In fact, comparison with Fig.~\ref{fig:abundance_x} shows that non--thermal
$\chi$ production leads to faster growth of $|\delta|$, and hence to earlier
and larger deviation between $Y_{1,r}$ and the exact solution of the Boltzmann
equation (\ref{eq:boltzmann_d}). However, comparison with the curves labeled
$Y_{\chi,{\rm tp}}$, where non--thermal $\chi$ production is neglected, show
that for this rather large cross section and short $\phi$ lifetime, the
non--thermal production mechanism does not affect the final $\chi$ relic
density any more. This agrees with the result of Fig.~\ref{fig:nonzero}, where
we saw that for the same values of $a$ and $x_0$, the relic density is
independent of the initial value $Y_\chi(x_0)$. As before, $Y_{1,r}$
approaches the exact result again for $x - x_0 \gg 1$. We therefore conclude
that our re--summed ansatz describes scenarios with additional non--thermal
$\chi$ production as well as the simpler case with only thermal production.

\section{Summary and Conclusions}

In this paper we investigated the relic abundance of non--relativistic
long--lived or stable particles $\chi$ using analytical as well as
numerical methods. Our emphasis was on scenarios with low re--heat
temperature, so that $\chi$ may never have been in full thermal equilibrium
after the end of inflation. Such scenarios are interesting because they lower
the predicted relic abundance and therefore open the parameter space of
particle physics models, allowing combinations of parameters which are
cosmologically disfavored in the standard high temperature scenario.

The case of small $\chi$ annihilation cross section or very low temperature
can easily be treated analytically, since in this case $\chi$ annihilation can
either be ignored completely, leading to our zeroth order solution $Y_0$ of
Eq.(\ref{eq:Y_0}), or can be treated as small perturbation, as in our first
order solution $Y_1$ of Eq.(\ref{eq:first_order}). Unfortunately this
approximation breaks down well before $\chi$ attains full thermal equilibrium.
On the other hand, we found that the simple trick of ``re--summing'' the
correction due to $\chi$ annihilation, as in Eq.(\ref{eq:resum}), allows to
describe the full temperature dependence of the $\chi$ number density as long
as $\chi$ does not reach full equilibrium. We saw in Sec.~4 that this remains
true even if a non--thermal source of $\chi$ production is added. Our ansatz
therefore provides a first analytical description of the ``in--between''
situation, where $\chi$ annihilation is very significant but not large enough
to establish full chemical equilibrium with the thermal plasma.

For yet higher cross sections or temperatures even the re--summed ansatz fails
to describe the temperature dependence of the $\chi$ number density at
intermediate temperatures. However, by replacing the initial scaled inverse
temperature $x_0$ with the quantity $x_{0,{\rm max}}$ of Eq.(\ref{eq:xmax})
our ansatz succeeds in predicting the final relic density about as well as
the standard semi--analytical high temperature treatment does, with comparable
numerical effort.

In this paper we have used the non--relativistic expansion of the $\chi$
annihilation cross section. This expansion is known to fail in certain cases
even for non--relativistic WIMPs \cite{except}. We expect our methods to be
applicable to these situations as well. However, a full analytical treatment
will be possible only if the product of thermally averaged cross section and
squared $\chi$ equilibrium number density, expressed as function of the scaled
inverse temperature $x$, can be integrated analytically over $x$.

From the particle physics point of view, the main effect of a low reheat
temperature is that it allows to reproduce the correct relic density in
scenarios with low annihilation cross section, e.g. for Bino--like neutralinos
and large sfermion masses. Conversely, the non--thermal production mechanism
studied in Sec.~4 allows to reproduce the correct relic density for WIMPs with
large annihilation cross section, e.g. Wino--like neutralinos \cite{moroi}. As
noticed in \cite{gg}, the combination of these effects in principle allows to
completely decouple the WIMP relic density from its annihilation cross
section. In many studies of expected WIMP detection rates scenarios yielding
too high a relic density under the standard assumptions were not considered;
such scenarios typically also lead to low detection rates. Conversely, in
scenarios leading to too low a thermal WIMP density, which typically predict
large detection rates for fixed WIMP density, the predicted detection rates
were often rescaled by the ratio of the predicted to the observed relic
density. If one allows lower reheat temperatures and/or non--thermal WIMP
sources the possible range of signals for WIMP detection can therefore be
enlarged towards both larger and smaller values.

In summary, we found analytical or semi--analytical solutions of the Boltzmann
equation describing the density of non--relativistic relics which are valid
for a wide range of initial conditions. In particular, they allow a complete
description of the temperature dependence for small or moderate cross
sections, and correctly reproduce the final relic density for {\em all}
combinations of initial temperature and cross section. This should be a
powerful tool for exploring the physics of non--relativistic relics,
especially in scenarios with low reheat temperature.

\subsection*{Acknowledgments}

The authors would like to thank the European Network of
Theoretical Astroparticle Physics ILIAS/N6 under contract
number RII3--CT--2004--506222 for financial support.
The work of M.K. is supported in part by the Japan Society for the Promotion
of Science.

\section*{Appendix}

In this Appendix, we give explicit expressions for the functions
$F^m_n(x,x_0)$, $G_n^r(x,x_0)$, $G_n^{r/2}(x,x_0)$ and $G_n^c(x,x_0)$ which
appear in Secs.~3 and 4. These functions are analytically expressed in terms
of the exponential integral of the first order ${\rm E}_1 (x)$ and the error
function ${\rm erfc}(x)$.

First we review the exponential integral and the error function.
The exponential integral of the first order is defined by
\begin{eqnarray} \label{ap1}
  {\rm E}_1 (x) = \int_1^\infty dt \ \frac{{\rm e}^{-xt}}{t}
  = \int_x^\infty dt \ \frac{{\rm e}^{-t}}{t}\, .
\end{eqnarray}
We need this function only for $x > x_0 \gg 1$. We can then use the asymptotic
large $x$ expansion,
\begin{eqnarray} \label{ap2}
{\rm E}_1 (x) \sim \frac{{\rm e}^{-x}}{x} \sum^\infty_{n=0} \frac{(-1)^n n!}{x^n}\, .
\end{eqnarray}
The error function is defined by
\begin{eqnarray} \label{ap3}
  {\rm erfc}(x) & = & \frac{2}{\sqrt{\pi}} \int_x^\infty dt \ {\rm e}^{-t^2} \, ,
\end{eqnarray}
with asymptotic large $x$ expansion
\begin{eqnarray} \label{ap4}
  {\rm erfc}(x) \sim \frac{{\rm e}^{-x^2}}{\sqrt{\pi}x}
  \sum^\infty_{n=0} \frac{(-1)^n (2n-1)!!}{(2x^2)^n}\, .
\end{eqnarray}

The functions $F^m_n(x,x_0)$ are defined by
\begin{eqnarray}
  F^m_n(x,x_0) = \int_{x_0}^x dt \ \frac{{\rm e}^{-mt}}{t^n}\, .
\end{eqnarray}
These integrals can be reduced to the form (\ref{ap1}). The resulting
expressions and corresponding asymptotic expansions, computed from
Eq.(\ref{ap2}), are:
\begin{eqnarray} \label{eq:app_f}
  F^4_0(x,x_0) & = & \frac{1}{4} ({\rm e}^{-4x_0} - {\rm e}^{-4x})\, , \nonumber \\
  F^4_1(x,x_0) & = & {\rm E}_1 (4 x_0) - {\rm E}_1 (4 x) \nonumber \\
  & \sim & \frac{{\rm e}^{-4x_0}}{4x_0} \left( 1 - \frac{1}{4 x_0} \right)
  - \frac{{\rm e}^{-4x}}{4x} \left( 1 - \frac{1}{4 x} \right)
  + {\cal O} \left(\frac{{\rm e}^{-4x_0}}{x_0^3} \right)\, , \nonumber \\
  F^4_2(x,x_0) & = & \frac{{\rm e}^{-4x_0}}{x_0} - 4 {\rm E}_1 (4x_0)
  - \frac{{\rm e}^{-4x}}{x} + 4 {\rm E}_1 (4x) \nonumber \\
  & \sim & \frac{{\rm e}^{-4x_0}}{4x_0^2} - \frac{{\rm e}^{-4x}}{4x^2}
  + {\cal O} \left(\frac{{\rm e}^{-4x_0}}{x_0^3} \right)\, , \nonumber \\
  F^4_3(x,x_0) & = & \frac{{\rm e}^{-4x_0}}{2x_0^2}
  - 2 \frac{{\rm e}^{-4x_0}}{x_0} + 8 {\rm E}_1(4 x_0)
  - \frac{{\rm e}^{-4x}}{2x^2}
  + 2 \frac{{\rm e}^{-4x}}{x} - 8 {\rm E}_1(4 x) \nonumber \\
  & \sim & {\cal O} \left(\frac{{\rm e}^{-4x_0}}{x_0^3} \right)\, , \nonumber \\
  F^2_1(x,x_0) & = & {\rm E}_1 (2 x_0) - {\rm E}_1 (2 x) \nonumber \\
  & \sim & \frac{{\rm e}^{-2x_0}}{2x_0} \left( 1 - \frac{1}{2 x_0} \right)
  - \frac{{\rm e}^{-2x}}{2x} \left( 1 - \frac{1}{2 x} \right)
  + {\cal O} \left(\frac{{\rm e}^{-2x_0}}{x_0^3} \right)\, , \nonumber \\
  F^2_2(x,x_0) & = & \frac{{\rm e}^{-2x_0}}{x_0} - 2 {\rm E}_1 (2x_0)
  - \frac{{\rm e}^{-2x}}{x} + 2 {\rm E}_1 (2x) \nonumber \\
  & \sim & \frac{{\rm e}^{-2x_0}}{2x_0^2} - \frac{{\rm e}^{-2x}}{2x^2}
  + {\cal O} \left(\frac{{\rm e}^{-2x_0}}{x_0^3} \right)\, , \nonumber \\
  F^2_3(x,x_0) & = & \frac{{\rm e}^{-2x_0}}{2 x_0^2}
  - \frac{{\rm e}^{-2 x_0}}{x_0} + 2 {\rm E}_1(2x_0)
  - \frac{{\rm e}^{-2x}}{2 x^2}
  + \frac{{\rm e}^{-2 x}}{x} - 2 {\rm E}_1(2x) \nonumber \\
  & \sim & {\cal O} \left(\frac{{\rm e}^{-2x_0}}{x_0^3} \right)\, , \nonumber \\
  F^0_2(x,x_0) & = & \frac{1}{x_0} - \frac{1}{x}\, , \nonumber \\
  F^0_3(x,x_0) & = & \frac{1}{2x_0^2} - \frac{1}{2x^2}\, .
\end{eqnarray}

The functions $G_n^r(x,x_0)$ and $G_n^{r/2}(x,x_0)$ are defined by
\begin{eqnarray}
G_n^r(x,x_0) &=& \int_{x_0}^x dt \ \frac{{\rm e}^{- r t^2}}{t^n}\, ,
  \quad n=2,3\, , \nonumber \\
G_n^{r/2}(x,x_0) &=& \int_{x_0}^x dt \ \frac{{\rm e}^{- r t^2/2}}{t^n}\, ,
  \quad n=2,3\, .
\end{eqnarray}
Using Eqs.(\ref{ap3}) and (\ref{ap4}), we find the following explicit
expressions and corresponding asymptotic expansions:
\begin{eqnarray} \label{eq:gi}
  G_2^r(x,x_0) &=& \frac{{\rm e}^{- r x_0^2}}{x_0}
  - \sqrt{\pi r} \ {\rm erfc}(\sqrt{r} x_0)
  - \frac{{\rm e}^{- r x^2}}{x}
  + \sqrt{\pi r} \ {\rm erfc}(\sqrt{r} x) \nonumber \\
  & \sim & \frac{{\rm e}^{-rx_0^2}}{2 r x_0^3} \left( 1 - \frac{3}{2 r x_0^2} \right)
  - \frac{{\rm e}^{-rx^2}}{2 r x^3} \left( 1 - \frac{3}{2 r x^2} \right)
  + {\cal O} \left( \frac{{\rm e}^{-r x_0^2}}{x_0 (rx_0^2)^3} \right)\, ,
  \nonumber \\
  G_3^r(x,x_0) &=& \frac{{\rm e}^{-r x_0^2}}{2 x_0^2}
  - \frac{r}{2} \ {\rm E}_1 (r x_0^2)
  - \frac{{\rm e}^{-r x^2}}{2 x^2}
  + \frac{r}{2} \ {\rm E}_1 (r x^2)
  \nonumber \\
  & \sim & \frac{{\rm e}^{-rx_0^2}}{2 r x_0^4} \left( 1 - \frac{2}{r x_0^2} \right)
  - \frac{{\rm e}^{-rx^2}}{2 r x^4} \left( 1 - \frac{2}{r x^2} \right)
  + {\cal O} \left( \frac{{\rm e}^{-r x_0^2}}{x_0^2 (rx_0^2)^3} \right)\, ,
  \nonumber \\
  G_2^{r/2}(x,x_0) &=& \frac{{\rm e}^{- r x_0^2/2}}{x_0}
  - \sqrt{\frac{\pi r}{2}} \ {\rm erfc}\left(\sqrt{\frac{r}{2}} x_0 \right)
  - \frac{{\rm e}^{- r x^2/2}}{x}
  + \sqrt{\frac{\pi r}{2}} \ {\rm erfc} \left(\sqrt{\frac{r}{2}} x \right)
  \nonumber \\ 
  & \sim & \frac{{\rm e}^{-rx_0^2/2}}{r x_0^3} \left( 1 - \frac{3}{r x_0^2} \right)
  - \frac{{\rm e}^{-rx^2/2}}{r x^3} \left( 1 - \frac{3}{r x^2} \right)
  + {\cal O} \left( \frac{{\rm e}^{-r x_0^2}}{x_0 (rx_0^2)^3} \right)\, ,
  \nonumber \\
  G_3^{r/2}(x,x_0) &=& \frac{{\rm e}^{-r x_0^2/2}}{2 x_0^2}
  - \frac{r}{4} \ {\rm E}_1 \left( \frac{r x_0^2}{2} \right)
  - \frac{{\rm e}^{-r x^2/2}}{2 x^2}
  + \frac{r}{4} \ {\rm E}_1 \left( \frac{r x^2}{2} \right)
  \nonumber \\
  & \sim & \frac{{\rm e}^{-rx_0^2/2}}{r x_0^4} \left( 1 - \frac{4}{r x_0^2} \right)
  - \frac{{\rm e}^{-rx^2/2}}{r x^4} \left( 1 - \frac{4}{r x^2} \right)
  + {\cal O} \left( \frac{{\rm e}^{-r x_0^2}}{x_0^2 (rx_0^2)^3} \right)\, .
\end{eqnarray}
In the expansion we assume that $rx_0^2 \sim x_0$, so that the effect of
non--thermal $\chi$ production is comparable to that of thermal production.

Finally, the functions $G_n^c(x,x_0)$ are defined by
\begin{eqnarray} \label{ap5}
  G_n^c(x,x_0) = \int_{x_0}^x dt \ \frac{{\rm e}^{-2t - r t^2/2}}{t^n}\, ,
  \quad n=1,2,3\, .
\end{eqnarray}
They appear in the ``interference terms'' in Eq.(\ref{eq:delta_d}), which are
important only if thermal and non--thermal contributions to $Y_0$ in
Eq.(\ref{eq:y0d}) are comparable in size. Since the overall $t-$dependence of
the integrand in Eq.(\ref{ap5}) is dominated by the numerator, we can, to good
approximation, evaluate these functions by replacing $t$ in the denominator by
some appropriate constant $x_c$:
\begin{eqnarray}
G_n^c(x,x_0) & \simeq & \int_{x_0}^x dt \ \frac{{\rm e}^{-2t - r t^2/2}}{x_c^n}
  \nonumber \\
  & = & \frac{{\rm e}^{2/r}}{x_c^n} \sqrt{\frac{\pi}{2r}}
  \left[ {\rm erfc} \left( \frac{1}{\sqrt{2r}} ( r x_0+ 2) \right) -
  {\rm erfc} \left(\frac{1}{\sqrt{2r}} (r x+ 2) \right)
  \right] \nonumber \\
  & \sim & \frac{{\rm e}^{-2x_0 - rx_0^2/2}}{x_c^n (r x_0 + 2)}
  \left[ 1 - \frac{r}{( r x_0 + 2)^2} \right]
  - \frac{{\rm e}^{-2x - rx^2/2}}{x_c^n (r x + 2)}
  \left[ 1 - \frac{r}{( r x + 2)^2} \right] \nonumber \\
  && + {\cal O}
  \left( \frac{{\rm e}^{-2x_0 - r x_0^2/2}}{x_0^{n-1}(rx_0^2)^3} \right)\, .
\end{eqnarray}
In our calculations in Sec.~5 we set $x_c = x_0$; this over--estimates $G_n^c$
by a few \%, with negligible error in $Y_{1,r}$.

\end{document}